\documentclass[10pt,final,twocolumn]{IEEEtran}

\usepackage{jhoagg}
\usepackage{cite}
\usepackage{amsmath}

\usepackage{dblfloatfix}
\usepackage{amsfonts}
\usepackage{amssymb,latexsym}
\usepackage[psamsfonts]{eucal}
\usepackage{amsthm}
\usepackage{graphicx}
\usepackage{epsfig}
\usepackage{psfrag}
\usepackage{subfigure}
\usepackage{indentfirst}
\usepackage{enumitem}
\usepackage{xcolor,pict2e}

\usepackage{array}
\newcolumntype{L}[1]{>{\raggedright\let\newline\\\arraybackslash\hspace{0pt}}p{#1}}
\newcolumntype{C}[1]{>{\centering\let\newline\\\arraybackslash\hspace{0pt}}p{#1}}
\newcolumntype{R}[1]{>{\raggedleft\let\newline\\\arraybackslash\hspace{0pt}}p{#1}}

\allowdisplaybreaks

\newtheorem{theorem}{\indent Theorem}

\newtheorem{proposition}{\indent Proposition}

\newtheorem{lemma}{\indent Lemma}

\newtheorem{corollary}{\indent Corollary}

\newtheorem{definition}{\indent Definition}

\newtheorem{example}{\indent Example}

\newtheorem{assumption}{\indent Assumption}

\newtheorem{algorithm}{\indent Algorithm}

\graphicspath{ {Figures/} }
\DeclareGraphicsExtensions{.png}

\usepackage{fancyhdr}

\fancypagestyle{firststyle}
{
    \fancyfoot{} 
   \fancyfoot[L]{\textit{Preprint submitted to IEEE Transactions on Cybernetics}}
}

\begin{document}

    \title{The Impact of Reference-Command Preview on Human-in-the-Loop Control Behavior}

    \author{
        Pedram~Rabiee,~
        S.~Alireza~Seyyed~Mousavi,~
        Amelia~J.~S.~Sheffler,~
        Erik~Hellstr\"{o}m,~
        Mrdjan~Jankovic,~\\
        Mario A. Santillo,~
        T.~M.~Seigler,~
        and~Jesse~B.~Hoagg

        \thanks{P. Rabiee, S. A. S. Mousavi, A. J. S. Sheffler, T. M. Seigler, and J. B. Hoagg are with the Department of Mechanical and Aerospace Engineering, University of Kentucky, Lexington, KY, USA. (e-mail: pedram.rabiee@uky.edu, amelia.sheffler@gmail.com, sse243@g.uky.edu, tmseigler@uky.edu, jesse.hoagg@uky.edu).}
        \thanks{E. Hellstr\"{o}m, M. Jankovic, and M. A. Santillo are with Research and Advanced Engineering, Ford Motor Company, Dearborn, MI, USA. (e-mail: jhells11@ford.com, Mrdjan@ieee.org, msantil3@ford.com).}
        \thanks{This work is supported in part by the Ford Motor Company and the National Science Foundation (1849213).}
    }

\maketitle
\thispagestyle{firststyle}

\begin{abstract}
This article presents results from an experiment in which 44 human subjects interact with a dynamic system to perform 40 trials of a command-following task.
The reference command is unpredictable and different on each trial, but all subjects have the same sequence of reference commands for the 40 trials. 
The subjects are divided into 4 groups of 11 subjects.
One group performs the command-following task without preview of the reference command, and the other 3 groups are given preview of the reference command for different time lengths into the future (0.5~s, 1~s, 1.5~s).
A subsystem identification algorithm is used to obtain best-fit models of each subject's control behavior on each trial.
The time- and frequency-domain performance, as well as the identified models of the control behavior for the 4 groups are examined to investigate the effects of reference-command preview.
The results suggest that preview tends to improve performance by allowing the subjects to compensate for sensory time delay and approximate the inverse dynamics in feedforward.
However, too much preview may decrease performance by degrading the ability to use the correct phase lead in feedforward.
\end{abstract}


\section{Introduction}
\label{sec:Introduction}

Command following is a common control problem in which the objective is to generate a control signal that makes the output of a dynamic system follow a reference command.
The reference command may be known only at the current instant of time, or it may be known for some time horizon into the future.
\textit{Preview} is knowledge of the reference-command trajectory into the future.

Driving a car on a winding road is an example of humans performing a command-following task with preview.
The objective in this case is to direct the car along the road, and the driver has preview because they have some knowledge of the path ahead.
This knowledge may be available because the path ahead is visible, the road is familiar, or there are road signs. 
Experience suggests that preview enables driving at higher speeds and is important for safety.
Some studies suggest that approximately one second of preview is needed for successful steering \cite{Land1995}.

It is unclear exactly how humans use preview for control; however, preview can allow for a broad range of control strategies that would be challenging to employ without preview.
For example, if the reference command and its derivatives (a type of preview) are available for feedback and the system being controlled is minimum phase (i.e., zeros in the open-left-half plane), then we can design a feedback controller to yield asymptotically perfect command following  \cite{hoagg2013a,hoagg2013b}. 
As another example, if the command is known in advance, then its frequency content can be used to design a feedback controller that relies on an internal model of the frequency content in the command \cite{young1972, davison1975b, francis1974, hoagg2008b}. 
Alternatively, feedforward inversion is a control strategy that uses the inverse system dynamics in feedforward.
Since the inverse system dynamics are typically improper, it follows that feedforward inversion may require preview for practical implementation. 
Furthermore, many control methods that optimize a prediction of system trajectories into the future (e.g., model-predictive control) rely on preview of the reference command.

Although the strategies that humans use to control dynamic systems are unknown, there are models that approximate human-in-the-loop (HITL) control behavior in certain scenarios \cite{Mulder2018}.
The \textit{crossover model} and the \textit{precision model} are linear time-invariant (LTI) controllers that approximate HITL compensatory (i.e., feedback) behavior in scenarios where only the command-following error is provided to the human \cite{mcruer1959, mcruer1959b, McRuer1965, mcruer1967, McRuer1967I, McRuer1967II, McRuer1968, McRuer1974}.
On the other hand, it is more challenging to model HITL control behavior for command following where humans have access to both feedback and feedforward.
Efforts to extend the crossover and precision models to address command-following behavior are presented in \cite{drop2013, laurense2015, Mulder2018b} for elementary dynamic systems, specifically, a static gain, single integrator, and double integrator.

Prior studies on command-following with preview include \cite{vanderEl2016, el2018l, vanderEl2018a}.
Specifically, \cite{vanderEl2016} develops an empirical HITL control model for interactions with elementary dynamic systems (static gain, single integrator, double integrator) based on data from HITL experiments with a combined command-following and disturbance-rejection objective. 
The model in \cite{vanderEl2016} is used in \cite{el2018l,vanderEl2018a} to numerically simulate the effects of preview on performance and control behavior, and the results are compared with HITL experiments.


A challenge of modeling HITL command-following behavior is that vastly different control strategies can yield similar qualitative features in the closed-loop response \cite{zhang2018a}.
Subsystem identification (SSID) methods have been developed to obtain the feedback and feedforward LTI models that are the best fit to the closed-loop data without an \textit{a priori} assumed control strategy \cite{zhang2016a, zhang2016b, seyyedmousavi2020a}. 
In \cite{zhang2018a, zhang2022a, seyyedmousavi2020b, seyyedmousavi2020c ,Koushkbaghi_JFI_Nonlinear, seyyedmousavi2018a}, SSID methods are applied to HITL experimental data to investigate command-following behavior.
The results in \cite{zhang2018a, zhang2022a, seyyedmousavi2020b, seyyedmousavi2020c, Koushkbaghi_JFI_Nonlinear, seyyedmousavi2018a} demonstrate that feedforward inversion is a predominant control strategy for a variety of dynamic systems if the command is predictable (which can thought of as virtual preview).
A human's ability to implement feedforward inversion may be diminished if the command is unpredictable \cite{seyyedmousavi2020c}; or the dynamic system has difficult characteristics such as nonminimum-phase zeros \cite{zhang2022a}, high relative degree \cite{seyyedmousavi2020b}, time delay \cite{seyyedmousavi2018a}, and nonlinearities \cite{Koushkbaghi_JFI_Nonlinear}.
In these cases, humans may employ alternate control strategies \cite{zhang2022a,seyyedmousavi2020c}.

This article contributes the literature on HITL control by providing new insights into the impact of preview on HITL behavior. 
We present results from HITL experiments in which 44 subjects interact with an LTI dynamic system 40 times over a one-week period. 
All subjects interact with the same third-order LTI dynamic system and have the same sequence of reference commands, which vary from trial-to-trial and are unpredictable. 
The subjects are divided into 4 groups, and each group has a different time length of preview. 
One group has no preview, while the other 3 groups are given a preview of 0.5~s, 1~s, and 1.5~s.
The time- and frequency-domain performance of the 4 groups are compared to investigate the effects of preview.
Next, the SSID algorithm in \cite{seyyedmousavi2020a} is used to identify an LTI model of each subject's control strategy.
The results are used to examine the impact of preview on the control strategies that the subjects learn.
Some preliminary results from this article appear in the conference article \cite{sheffler2019a}; however, \cite{sheffler2019a} only includes groups with 0-s and 1-s preview.
This article also presents analyses that go significantly beyond the preliminary conference publication \cite{sheffler2019a}.

\section{Methods}\label{sec:Experimental Methods}

Forty-four people voluntarily participated in this study. 
At the time of the study, the subjects had no known neurological or motor control disorders and were 18 to 35 years of age. 
The University of Kentucky Institutional Review Board approved this study under IRB number 44649.

\subsection{Experiment}

\begin{figure}[t!]
\center{\includegraphics[width=0.49\textwidth,clip=true,trim= 0in 0in 0in 0in] {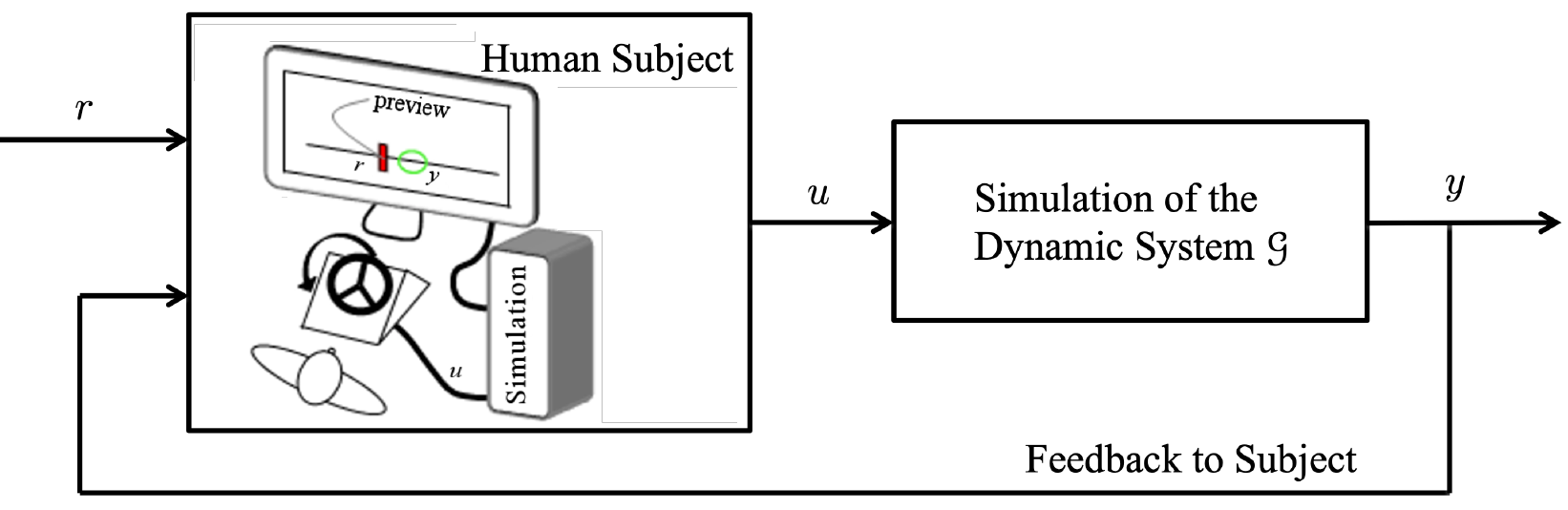}}
\caption{
Subjects use a steering wheel to affect the motion of a controlled object (green circle) on a computer screen.
The object's position $y$ is the output of a dynamic system, and the steering wheel position $u$ is the input to the dynamic system.
A reference object (red line) also moves on the computer screen, and its horizontal position is denoted by $r$.
}\label{fig:Figure1an}
\end{figure}

Subjects use a steering wheel to affect the horizontal position of a controlled object that is displayed on a computer screen.
The position of the steering wheel is denoted by $u$, which is the input to an LTI dynamic system. 
The horizontal position of the controlled object is denoted by $y$, which is the output of the LTI dynamic system. 
A reference object also moves on the computer screen, and its horizontal position is denoted by $r$. 
The position $r$ of the reference object is independent of $u$.
The signals $u$, $y$, and $r$ are functions of time~$t$.
Figure~\ref{fig:Figure1an} is a diagram of the experimental setup.

The controlled object's position $y$ satisfies the LTI differential equation
\begin{equation}
\dddot{y}(t) + 5.2\ddot{y}(t) + 9.76\dot{y}(t) + 6.4 y(t) = 3.2\dot{u}(t) + 7.04u(t).
\label{eq:LTIeq}
\end{equation}  
Hence, the transfer function from $u$ to $y$ is
\begin{equation*}
\SG(s) \triangleq \frac{3.2(s+2.2)}{(s+1.6)(s^2+3.6s+4)},
\end{equation*}
which has poles at $-1.6$ and $ -1.8 \pm \jmath 0.87$, and a zero at $-2.2$.

Prior to interacting with the experimental setup, each subject is shown the computer screen and told that manipulating the steering wheel moves the controlled object. 
Subjects are told that their objective is to manipulate the steering wheel and attempt to make the controlled and reference objects have the same horizontal position at each instant of time. 
Thus, each subject's objective is to generate a control $u$ that makes the magnitude of the error $e \triangleq r-y$ as small as possible. 
The subjects have no prior knowledge of the reference object's trajectory, or the dynamics \eqref{eq:LTIeq} that relate $u$ and $y$.

A trial is a $60$-s period during which a subject manipulates the steering wheel. 
Each subject performed $40$~trials of the experiment over $7$~days. 
These trials were divided into 4~sessions of 10~trials, and each session was completed in a $20$-min period.
Each subject completed no more than one session in a $12$-h period.
For each session, a subject sits in a chair facing the computer screen, which is approximately $60$~cm from the subject's eyes. 
The subject uses both hands to manipulate a 7-inch-diameter rotational steering wheel, which is connected to a rotational sensor (Groovy Game Gear model TRBOTWST201) that provides measurement $u(t)$. 
Then, the LTI differential equation \eqref{eq:LTIeq} is solved in real-time using custom software to calculate $y(t)$, which determines the position of the controlled object on the computer screen.

For $i \in  \{1, 2, \ldots , 40\}$, consider the reference-command signal
\begin{equation*}
c_{i}(t) \triangleq \frac{3}{10} \sum_{j=1}^{30} \cos\bigg( \frac{\pi jt}{30} + \phi_{i,j} \bigg) ,
\end{equation*}
where $\phi_{i,1},\phi_{i,2},\ldots,\phi_{i,30} \in [0,2 \pi)$ are randomly selected phases such that $c_i(0) = 0$ and the peak magnitude is less than $2.6$, that is, $\max_{t \in [0,60]} |c_i(t)| < 2.6$. 
Thus, $c_i$ is a $60$-s sum of $30$ sinusoids with evenly spaced frequencies between $0$ and $0.5$~Hz and with randomly selected phases $\phi_{i,1},\ldots,\phi_{i,30}$.

For each of the 44 subjects and for each trial $i \in \{1,2,\ldots,40 \}$, the reference object's position is $r(t)=c_i(t)$.
Therefore, the reference command is different on each trial, but each subject has the same sequence of reference commands $c_1,c_2,\ldots,c_{40}$ for the 40 trials.
We note that the reference object's trajectory $r$ is unpredictable because the phases $\phi_{i,1},\ldots,\phi_{i,30}$ are different on each trial.

The units of the reference $r$ are hash marks (hm), which indicates the horizontal position of the reference object on the computer screen. 
Each hash mark corresponds to a distance of 5.45~cm on the computer screen.
Note that $0$~hm denotes the center of the computer screen. 
The peak magnitude of $r$ is no more than $2.6$~hm, and the range of motion displayed on the computer screen is $\pm 4.4$~hm.

\subsection{Effect of Preview}

\begin{figure}[t!]
\center{
\includegraphics[width=0.47\textwidth,clip=true,trim= 0.8in 3.4in 0.1in 1.9in] 
{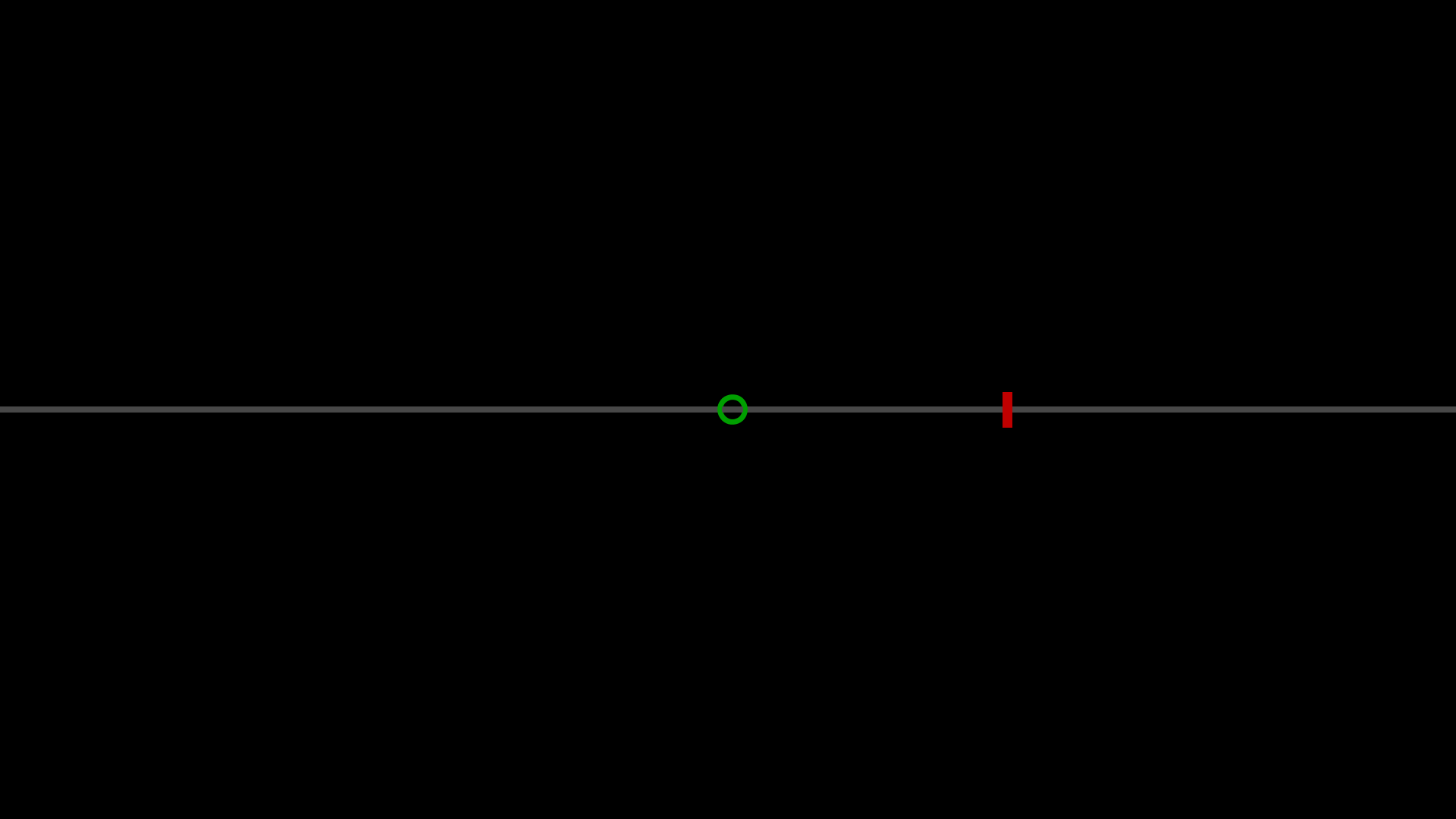}}
\caption{
Display for group~1 does not include preview of the reference command. 
The red line is the reference object's position and the green circle is the controlled object's position.}
\label{fig:0sPreviewScreenshot}
\end{figure}

\begin{figure}[t!]
\center{
\includegraphics[width=0.47\textwidth,clip=true,trim= 0.8in 3.4in 0.1in 1.9in] 
{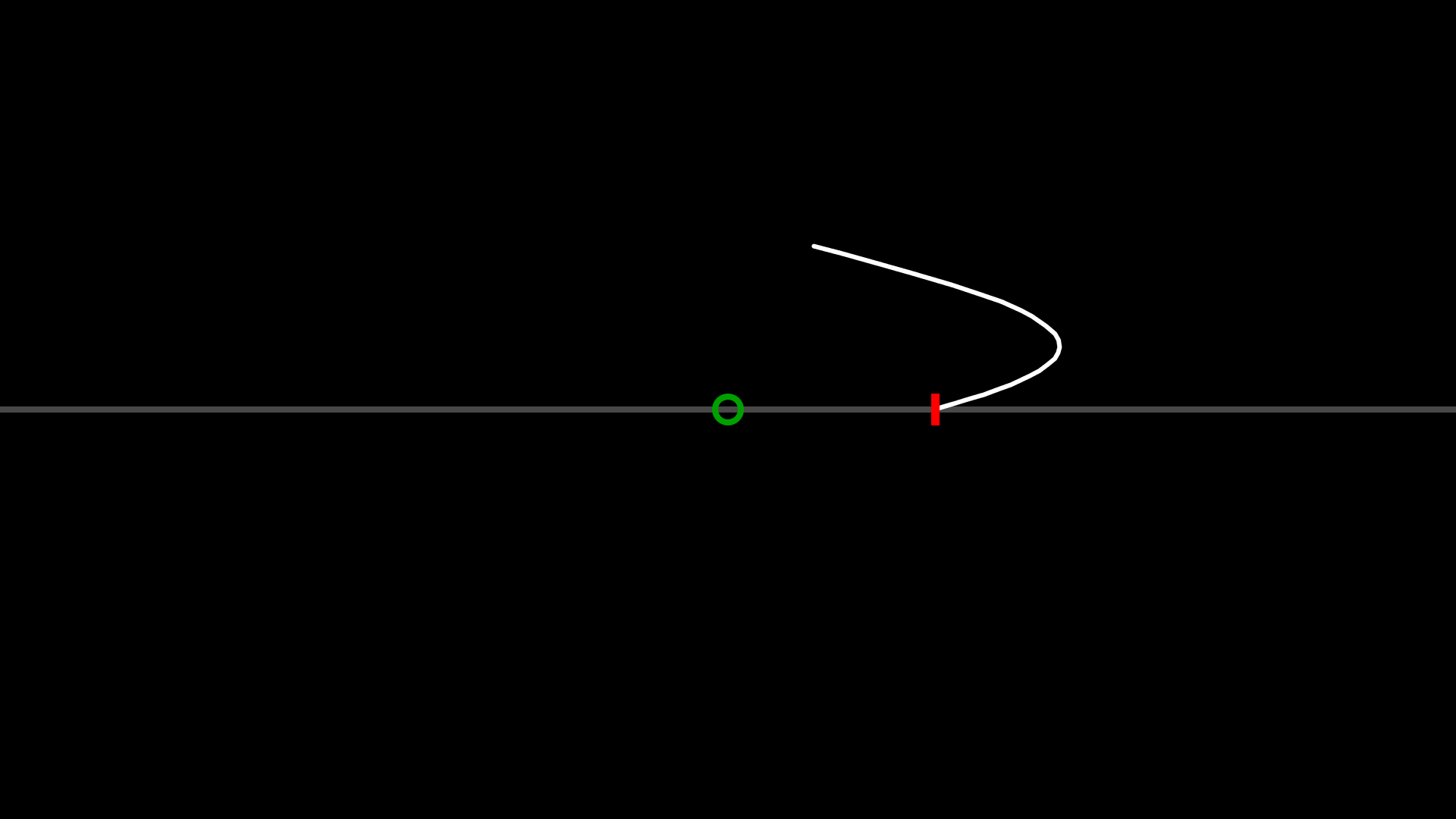}}
\caption{
Display for group~2 includes $1$-s preview of the reference command. 
Preview is displayed as a white curve above the reference object, which is the red line. 
The green circle is the controlled object's position.}
\label{fig:1sPreviewScreenshot}
\end{figure}

To examine the effects of reference-command preview, we divide the 44 subjects into 4 groups, where each group has 11 subjects.
For group~1, subjects are presented with a command-following (i.e., pursuit) display without preview.
Specifically, the current reference object's position $r(t)$ is displayed as a red line, and the current controlled object's position $y(t)$ is displayed as a green circle; both of which move horizontally on the computer screen. 
The command-following display without preview is shown in Fig.~\ref{fig:0sPreviewScreenshot}.
For groups~2--4, subjects are presented with a command-following display with preview. 
Similar to group~1, the current reference object's position $r(t)$ is displayed as a red line, and the current controlled object's position $y(t)$ is displayed as a green circle.
In addition, the trajectory of the reference object for some time into the future is also displayed on the computer screen.
Specifically, the screen shows $0.5$~s, $1$~s, and $1.5$~s of preview for groups 2, 3, and 3, respectively. 
This preview of the reference command is displayed as a white curve above the red line. 
As time progresses, the future trajectory of the reference object scrolls down the screen, and the current reference object's position $r(t)$ takes the position indicated by the future trajectory displayed above. 
For example, the display for group~2 allows subjects to preview the reference command 1~s into the future, that is, see the trajectory that the reference object will follow at all times from 0~s to 1~s into the future.
The command-following display with preview is shown in Fig.~\ref{fig:1sPreviewScreenshot}.

For each trial of each subject, we record $r$, $u$, and $y$ with a sample time of $T_\rms = 0.02$~s, which yields $n= 3000$ samples.
The sampled data obtained from $r$, $u$, and $y$ are denoted by $\{ r_k \}_{k=1}^n$, $\{ u_k \}_{k=1}^n$, and $\{ y_k \}_{k=1}^n$. 
For $k \in \{1,\ldots,n \}$, define $e_k \triangleq r_k - y_k$, which is the command-following error.

A \textit{divergent trial} is a trial, where for any $k \in \{ 1,\ldots,n \}$, $y_k$ exceeds $\pm 4.4$~hm display limits.
As shown in Table~\ref{table:divergent}, there are more divergent trials during the earlier trials than during the later trials.
Group~1 (no preview) has the most divergent trials---a total of $68$, which is $15\%$ of the trials. 
Group~3 ($1$-s preview) has the least divergent trials---a total of $18$, which is $4.1\%$ of the trials.
Divergent trials are omitted from the results reported in the rest of this paper.

\begin{table}[t]
	\centering
	\caption{Number of divergent trials.}
	\label{table:divergent}
		\begin{tabular}
	{C{0.3in} | C{0.26in} C{0.26in} C{0.28in} C{0.28in} C{0.28in} C{0.28in} | C{0.23in}}
    & Trials &  Trials  & Trials & Trials & Trials & Trials &\\
	Group & 1--5 &  6--10  & 11--20 & 21--30 & 31--35 & 36--40 &Total\\
		\hline
		1       & $18$ & $15$ & $14$ & $13$ & $5$ & $3$ & $68$\\
		2      & $15$ & $10$ & $13$ & $5$ & $4$ & $0$ & $47$\\
		3       & $9$ & $3$ & $4$ & $0$ & $2$ & $0$ & $18$\\
		4     & $24$ & $14$ & $7$ & $6$ & $1$ & $1$ & $53$\\
	\end{tabular}
\end{table}

\subsection{Modeling Control Strategies Using SSID}
\label{sec:SSID}

\begin{figure}[t]
	\centering
	\scalebox{1}{
		\setlength{\unitlength}{.008in}
		
		\begin{picture}(500,200)
		\thicklines
		
		\put(0,130){\vector(1,0){140}}
		\put(10,137){$r_k$}
		
		\put(0,50){\vector(1,0){70}}
		\put(10,57){$y_k$}
		
		\put(110,57){$e_k$}
		
		\put(80,130){\vector(0,-1){70}}

		\put(80,50){\circle{20}}
		\put(85,65){\line(1,0){10}}
		\put(90,60){\line(0,1){10}}
		\put(60,35){\line(1,0){10}}
		
		\put(90,50){\vector(1,0){50}}

		\put(40,10){\dashbox(330,190)}
		\put(50,180){Model of Subject's Control Strategy}
		
		\put(140,100){\framebox(60,60){$\begin{array}{c} \mbox{Delay} \\ z^{-\tau_{\rm ff}} \end{array}$}}
		\put(200,130){\vector(1,0){25}}

		\put(225,100){\framebox(95,60){$\begin{array}{c} \mbox{Feedforward} \\ G_{\rm ff} \end{array}$}}
		\put(320,130){\line(1,0){30}}
		\put(350,130){\vector(0,-1){30}}
		
		\put(140,20){\framebox(60,60){$\begin{array}{c} \mbox{Delay} \\ z^{-\tau_{\rm fb}} \end{array}$}}
		\put(200,50){\vector(1,0){40}}
		
		\put(240,20){\framebox(80,60){$\begin{array}{c} \mbox{Feedback} \\ G_{\rm fb} \end{array}$}}
		\put(320,50){\line(1,0){30}}
		\put(350,50){\vector(0,1){30}}
		
		\put(350,90){\circle{20}} 
		\put(345,90){\line(1,0){10}}
		\put(350,85){\line(0,1){10}}
		\put(360,90){\vector(1,0){50}}
		\put(405,97){$u_k$}
		
		\end{picture}}
	\caption{The control strategy is modeled using feedforward transfer function $G_{\rm ff}$, feedforward delay $\tau_{\rm ff}$, feedback transfer function $G_{\rm fb}$, and feedback delay $\tau_{\rm fb}$.}
	\label{fig:ControlStrategy}
\end{figure}
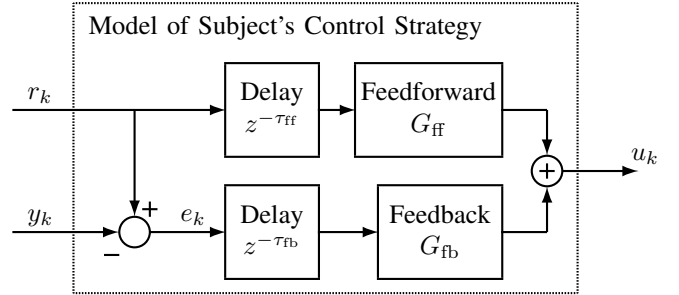

Each subject's control strategy on each trial is modeled by the LTI control shown in Fig.~\ref{fig:ControlStrategy}, which is given by
\begin{equation}
	\hat u(z) = z^{-\tau_{\rm fb}} G_{\rm fb}(z) \hat e(z) + z^{-\tau_{\rm ff}}G_{\rm ff}(z) \hat r(z), \label{eq:ControlArchitecture}
\end{equation}
where $\hat r(z)$, $\hat e(z)$, and $\hat u(z)$ are the $z$-transforms of $r_k$, $e_k$, and $u_k$; $G_{\rm fb}$ and $G_{\rm ff}$ are the transfer functions of the feedback and feedforward controllers; and the nonnegative integers $\tau_{\rm fb}$ and $\tau_{\rm ff}$ are the feedback and feedforward delays. 
Feedforward is the anticipatory control determined solely from the reference $r_k$, whereas feedback is the reactive control determined from the observed error $e_k$.
Define $T_{\rm fb} \triangleq 10^{3} \tau_{\rm fb} T_\rms$ and $T_{\rm ff} \triangleq 10^{3} \tau_{\rm ff} T_\rms$, which are the feedback and feedforward time delays in milliseconds.

Next, let $G$ be the discrete-time transfer function obtained by discretizing $\SG$ using a zero-order hold on the input with sample time $T_{\rm s}=0.02$ s. 
The closed-loop transfer function from $r_k$ to $y_k$ is
\begin{equation}
    \tilde G_{yr}(z) \triangleq  \frac{G(z) \left [ z^{-\tau_{\rm ff}}G_{\rm ff}(z) + z^{-\tau_{\rm fb}} G_{\rm fb}(z) \right ]}{1+ z^{-\tau_{\rm fb}} G_{\rm fb}(z) G(z)}. \label{eq:tildeGyr}
\end{equation}

For each trial, we calculate the discrete Fourier transform (DFT) of $\{y_k\}_{k=1}^n$ and $\{r_k\}_{k=1}^n$ at the frequencies $\omega_i = 2\pi i/60 $ rad/s, where $i \in \{1, 2, \ldots,N \}$, which are $N = 30$ evenly spaced frequencies over the $0$-to-$0.5$~Hz range of the reference command. 
Let $y_{ \rm dft }(\omega_i)$ and $r_{ \rm dft }(\omega_i)$ denote the DFT of $\{y_k\}_{k=1}^n$ and $\{r_k\}_{k=1}^n$ at $\omega_i$.
For each trial and for $i \in \{1, 2, \dots, N\}$, define $H(\omega_i) \triangleq {y_{\rm dft}( \omega_i)}/{r_{\rm dft}( \omega_i)}$, which is the closed-loop frequency-response data from $r_k$ to $y_k$.

For each trial, we use the SSID algorithm in \cite{seyyedmousavi2020a} to determine the control strategy (i.e., $G_{\rm ff}$, $\tau_{\rm ff}$, $G_{\rm fb}$, $\tau_{\rm fb}$) of the form \eqref{eq:ControlArchitecture} such that the modeled frequency response $\tilde{G}_{yr}(e^{\jmath\omega_i T_{\rm s}})$ is the best fit to the frequency-response data $ H(\omega_i)$.
Specifically, we seek to find $G_{\rm ff}$, $\tau_{\rm ff}$, $G_{\rm fb}$, and $\tau_{\rm fb}$ that minimizes the cost function
\begin{equation}
J( G_{\rm ff} ,  \tau_{\rm ff}, G_{\rm fb}, \tau_{\rm fb}) \triangleq  \sum_{i=1}^{N} \left | \tilde{G}_{yr}(e^{\jmath\omega_i T_{\rm s}}) - H(\omega_i) \right |^2
\label{eq:Cost}
\end{equation}
subject to the constraint that $\tilde{G}_{yr}$ is asymptotically stable. 
The orders of $G_{\rm fb}$ and $G_{\rm ff}$ are selected to allow for a range of control behaviors.
In this work, $G_{\rm ff}$ is second order, exactly proper, and finite impulse response (FIR); and $G_{\rm fb}$ is second order and strictly proper. 
The assumption that $G_{\rm ff}$ is FIR reduces computational complexity of the SSID algorithm but does not impose significant restriction on the type of feedforward behavior that can be modeled \cite{zhang2016b}.

The SSID method in \cite{seyyedmousavi2020a} is summarized as follows.
First, we generate 2 candidate pools. 
The feedback candidate pool contains possible models of $G_{\rm fb}$ and $\tau_{\rm fb}$. 
Every element in the feedback candidate pool is such that $\tilde G_{yr}$ is asymptotically stable. 
The feedforward-delay candidate pool contains possible values of $\tau_{\rm ff}$.
For each possible model in the candidate pools, the cost $J$ is convex in the coefficients of $G_{\rm ff}$. 
Thus, for each model in the feedback candidate pool, we solve a sequence of convex optimizations to find the best-fit $G_{\rm ff}$ and $\tau_{\rm ff}$.
Then, we search the feedback candidate pool to determine the quadruple $(G_{\rm ff}, \tau_{\rm ff}, G_{\rm fb}, \tau_{\rm fb})$ that minimizes $J$.
The candidate pools used in this work are the same as those in \cite[Appendix B]{seyyedmousavi2020c}.

The closed-loop frequency-response data $\{ H(\omega_i) \}_{i=1}^N$ from the HITL experiments in this work does not have significant variation in magnitude over the 0-to-0.5~Hz range. 
Thus, we use the cost \eqref{eq:Cost}, which has equal weight at each frequency $\omega_1,\ldots,\omega_N$.
If the closed-loop frequency-response data has significant magnitude variation, then it can be beneficial to weight each term in \eqref{eq:Cost} in order to normalize the magnitudes (e.g., weight each term by $| H(\omega_i) |^{-1}$). 
Normalization helps prevents the SSID from being biased toward minimizing only the terms where $|H(\omega_i) |$ is large. 
See \cite{seyyedmousavi2020a} for details on weighting.

\section{Effect of Preview on Performance} 
\label{sec:Experimental Data}

\begin{figure}[t!]
\center{
\includegraphics[width=0.48\textwidth,clip=true,trim= 0.0in 0.2in 0.7in 0.2in] 
{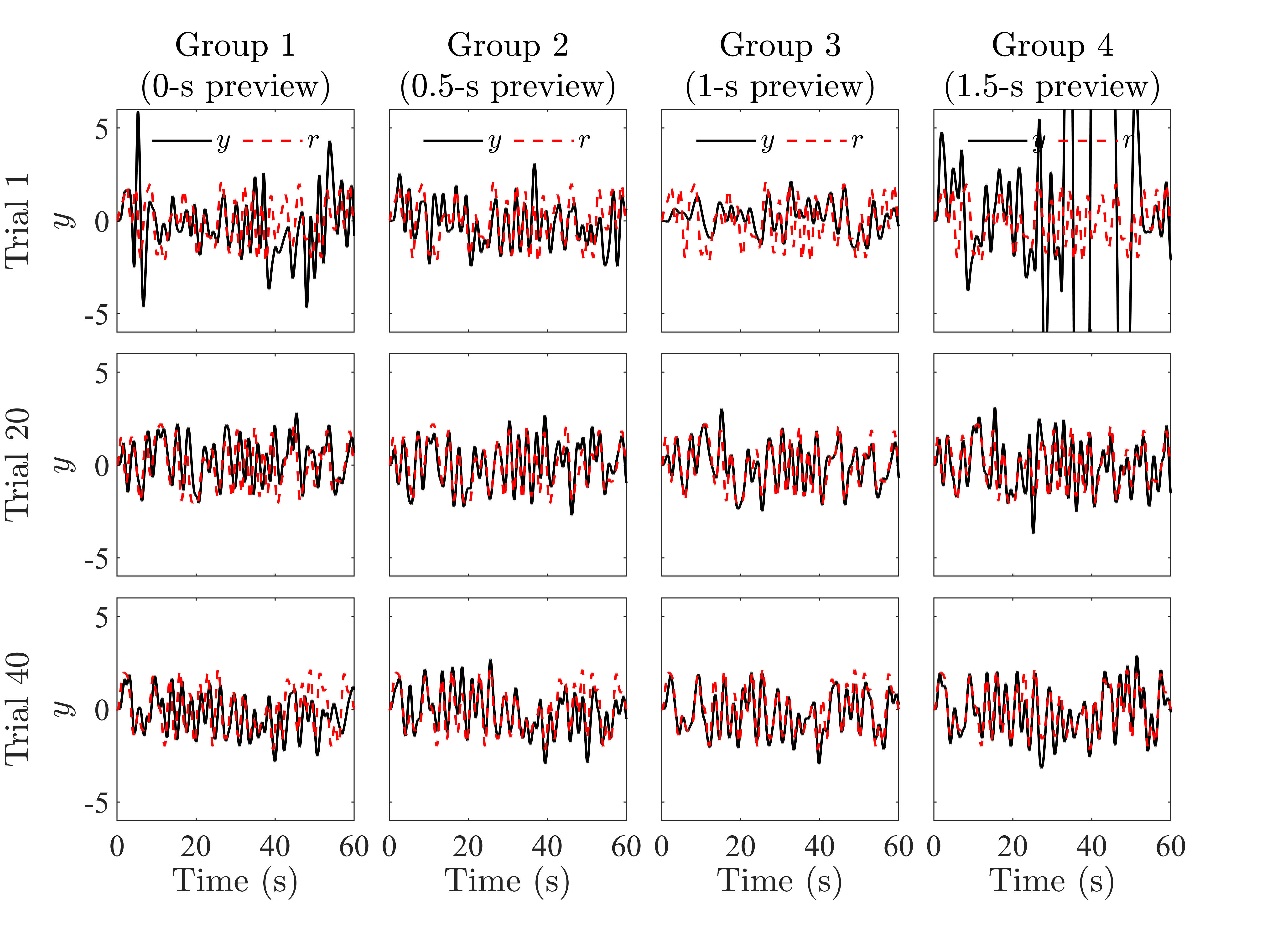}}
\caption{
Output $y$ and reference $r$ on trials 1, 20, and 40 of the subject from each group whose $\|e\|$ on the last trial is the median of the group.}
\label{fig:r_y_median}
\end{figure}

\begin{figure}[t!]
\center{
\includegraphics[width=0.48\textwidth,clip=true,trim= 0.0in 0.2in 0.7in 0.2in] 
{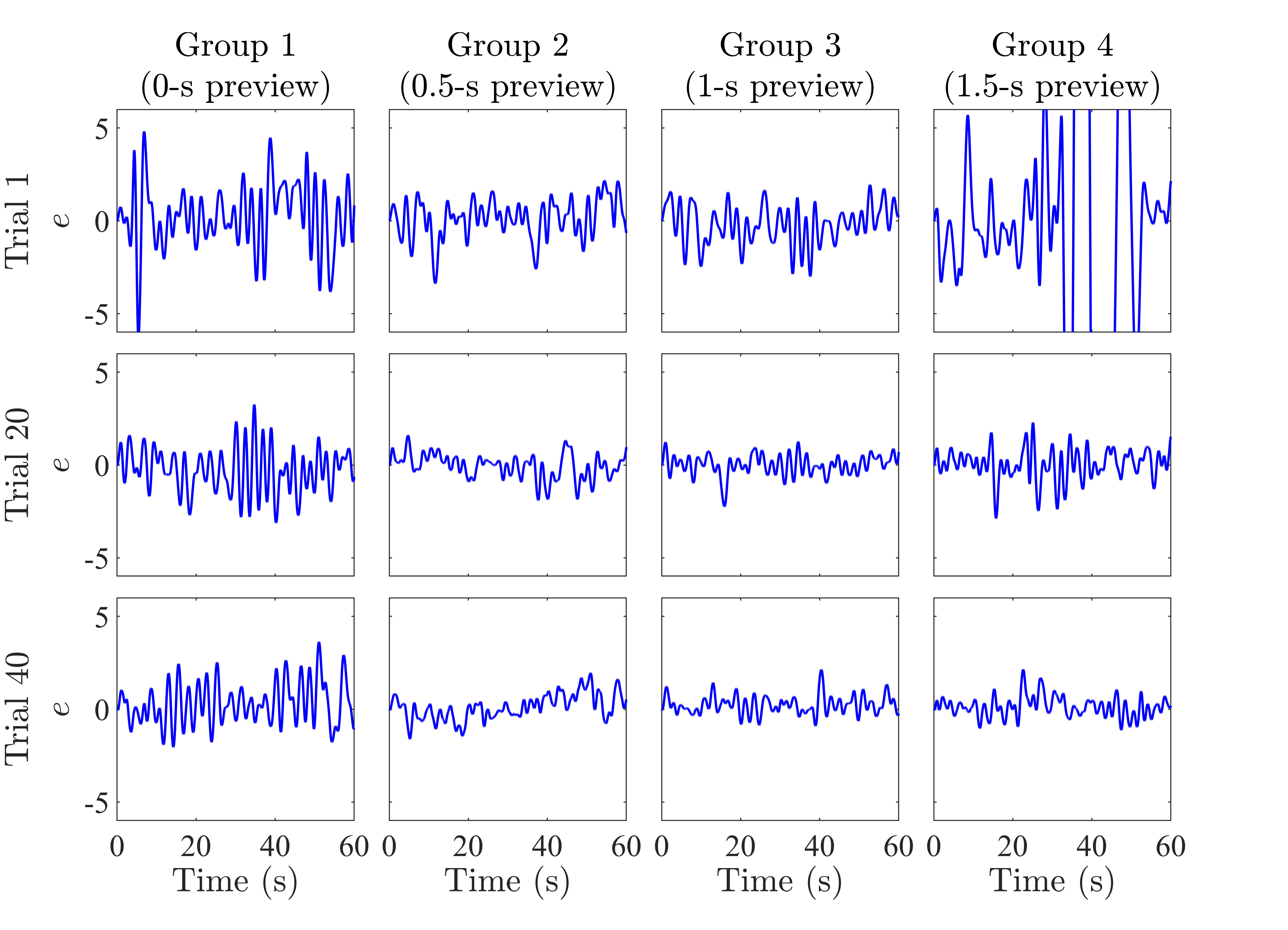}}
\caption{
Error $e$ on trials 1, 20, and 40 of the subject from each group whose $\|e\|$ on the last trial is the median of the group.}
\label{fig:error_median}
\end{figure}

For each trial, the time-averaged error is
\begin{equation*}
\|e\| \triangleq \frac{1}{n} \sum_{k=1}^n  |e_k |.
\end{equation*}
Figures~\ref{fig:r_y_median} and~\ref{fig:error_median} show $y$, $r$ and $e$ for trials~1, 20, and 40 of the subject from each group whose $\|e\|$ on the last trial is the median (i.e., 6th best) of the subjects in the group.
For each group, the median subject's $\|e\|$ on the last trial is less than their $\|e\|$ on the first trial. 
For each group, the median subject's command-following performance improves from trial~1 to trial~40. 
The time-averaged error on the last trial for the subject in group~1 is greater than that for the subject in group~2, which is greater than that for the subject in group~4, which is greater than that for the subject in group~3.


\begin{figure}[t!]
\center{
\includegraphics[width=0.48\textwidth,clip=true,trim= 0.2in 0.1in 0.75in 0.46in] 
{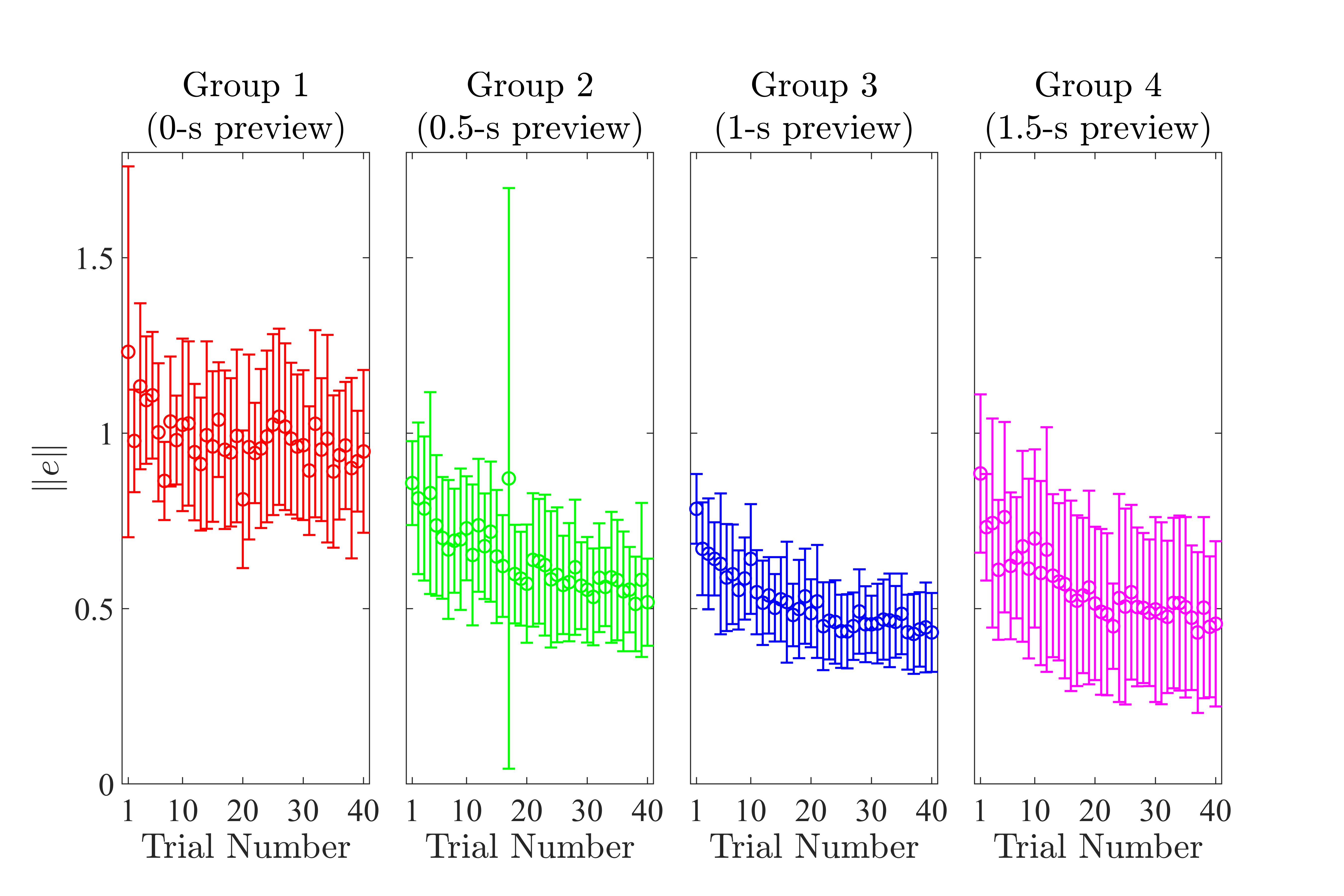}}
\caption{
Mean and standard deviation of $\|e\|$ on each trial for each group. 
The $\circ$ is the mean, and the lines indicate the standard deviation.
}\label{fig:error}
\end{figure}

Figure~\ref{fig:error} shows the mean and standard deviation of $\| e\|$ on each trial for each group, and Table~\ref{table:error} show mean $\| e\|$ for each group on different sets of trials.
For each group, mean $\| e\|$ tends to decrease from trial~1 to~40. 
By trial~40, mean $\|e\|$ is $6\%$, $49\%$, $57\%$, and $54\%$ better than the no-control (i.e, $u = 0$) response for groups~1--4, respectively.
The no-control (i.e, $u = 0$) time-averaged error is at least $0.96$ for each trial. 
Table~\ref{table:error} shows that  group~1 has little improvement over the no-control response. 
We also note that mean $\|e\|$ does not change significantly between trials~21 and~40, suggesting that subjects reach near-steady performance.

For each trial, mean $\|e\|$ for group~1 is greater than that of the other groups.
A one-way ANOVA comparing each group's mean $\|e\|$ over the last $5$ trials yields $F_{3,40} = 22.6$ and $p< 0.001$, thus confirming a statistical difference between groups. 
A Tukey post-hoc pairwise comparison yields $p<0.001$ for pairs $(1,2)$, $(1,3)$, and $(1,4)$. 
The $p$-values for pairs $(2,3)$, $(2,4)$, and $(3,4)$ are $p=0.43$, $p=0.66$, and $p=0.98$.
Since the task changes on each trial, the subjects have limited ability to predict the reference. 
Thus, the difference in mean $\| e \|$ between group~1 and the other groups indicates that preview helps to improve the performance.

\begin{table}[t!]
	\centering
	\caption{Mean $\| e\| $}
	\label{table:error}
	\begin{tabular}
	{C{0.29in} | C{0.26in} C{0.26in} C{0.28in} C{0.28in} C{0.28in} C{0.28in}}
     & Trials &  Trials  & Trials & Trials & Trials & Trials \\
    Group & 1--5 &  6--10  & 11--20 & 21--30 & 31--35 & 36--40 \\
		\hline
		1       & $1.11$ & $0.98$ & $0.96$ & $0.98$ & $0.95$ & $0.93$ \\
		2       & $0.80$ & $0.70$ & $0.67$ & $0.59$ & $0.57$ & $0.54$ \\
		3       & $0.67$ & $0.59$ & $0.51$ & $0.46$ & $0.47$ & $0.44$ \\
		4       & $0.74$ & $0.65$ & $0.57$ & $0.50$ & $0.50$ & $0.46$ \\
	\end{tabular}
\end{table}

For each set of trials in Table~\ref{table:error}, mean $\|e\|$ for group~3 is less than that of the other groups (although mean $\|e\|$ for groups~3 and~4 are close of the last 20 trials). 
Since group~3 has the lowest mean $\|e\|$ across trials, it is possible that there is an optimal amount of preview, which would depend on the dynamic system as well as the frequency content of the command. 
In other words, it may be possible to have either too little or too much preview.

Next, we examine frequency-domain performance.
For each trial, the \textit{frequency-averaged error in the magnitude of $y$} is
\begin{align*}
E_{\rm m} &\triangleq \frac{1}{N} \sum_{i=1}^{N} \bigg | \left | y_{\rm dft}(\omega_i) \right |e^{\jmath \angle r_{\rm dft}(\omega_i)} - \left | r_{\rm dft}(\omega_i) \right | e^{\jmath \angle r_{\rm dft}(\omega_i)} \bigg |\\
&= \frac{1}{N} \sum_{i=1}^{N} \bigg | \left | y_{\rm dft}(\omega_i) \right | - \left | r_{\rm dft}(\omega_i) \right | \bigg |,
\end{align*}
and the \textit{frequency-averaged error in the phase of $y$} is
\begin{align*}
E_{\rm p} &\triangleq \frac{1}{N} \sum_{i=1}^{N} \bigg | \left | r_{\rm dft}(\omega_i) \right |e^{\jmath \angle y_{\rm dft}(\omega_i)} - \left | r_{\rm dft}(\omega_i) \right | e^{\jmath \angle r_{\rm dft}(\omega_i)} \bigg |\\
&= \frac{1}{N} \sum_{i=1}^{N} \left | r_{\rm dft}(\omega_i) \right | \left | e^{\jmath \angle y_{\rm dft}(\omega_i)} - e^{\jmath \angle r_{\rm dft}(\omega_i)} \right |,
\end{align*}
where $\angle z$ is the angle of the complex number $z$.
Note that $E_{\rm m}$ is the frequency-averaged magnitude of the difference between the $y_{\rm dft}$ and $r_{\rm dft}$ assuming that the phase of $y_{\rm dft}$ is equal to the phase of $r_{\rm dft}$.
In contrast, $E_{\rm p}$ is the frequency-averaged magnitude of the difference between the $y_{\rm dft}$ and $r_{\rm dft}$ assuming that the magnitude of $y_{\rm dft}$ is equal to the magnitude of $r_{\rm dft}$.

\begin{figure}[ht!]
\center{\includegraphics[width=0.48\textwidth,clip=true,trim= 0.2in 0.1in 0.75in 0.46in] 
	{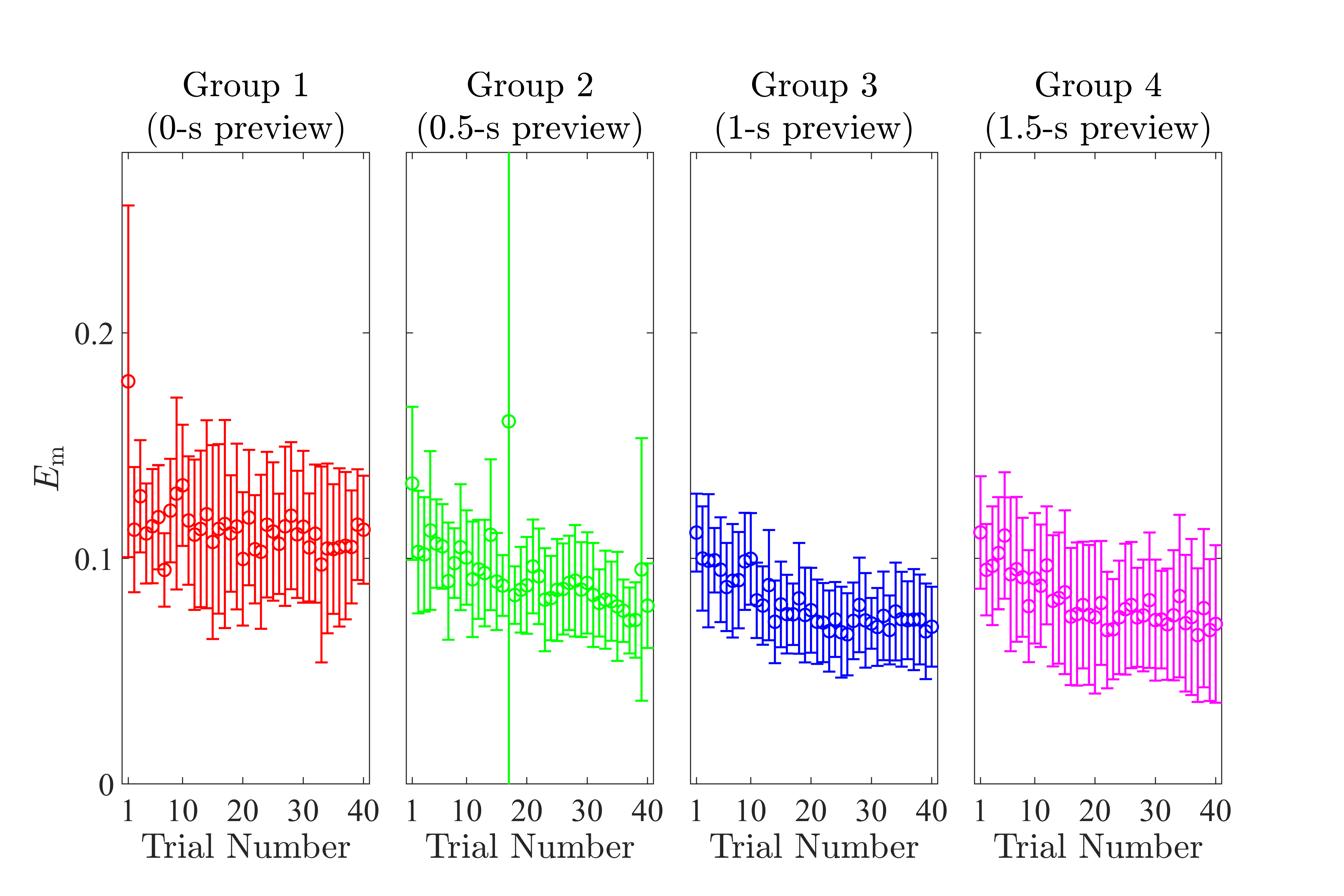}}
	\caption{
		Mean and standard deviation of $E_{\rm m}$ on each trial for each group.
		The $\circ$ is the mean, and the lines indicate the standard deviation.
		}\label{fig:Em}
\end{figure}

\begin{figure}[ht!]
	\center{\includegraphics[width=0.48\textwidth,clip=true,trim= 0.2in 0.1in 0.75in 0.48in]
	{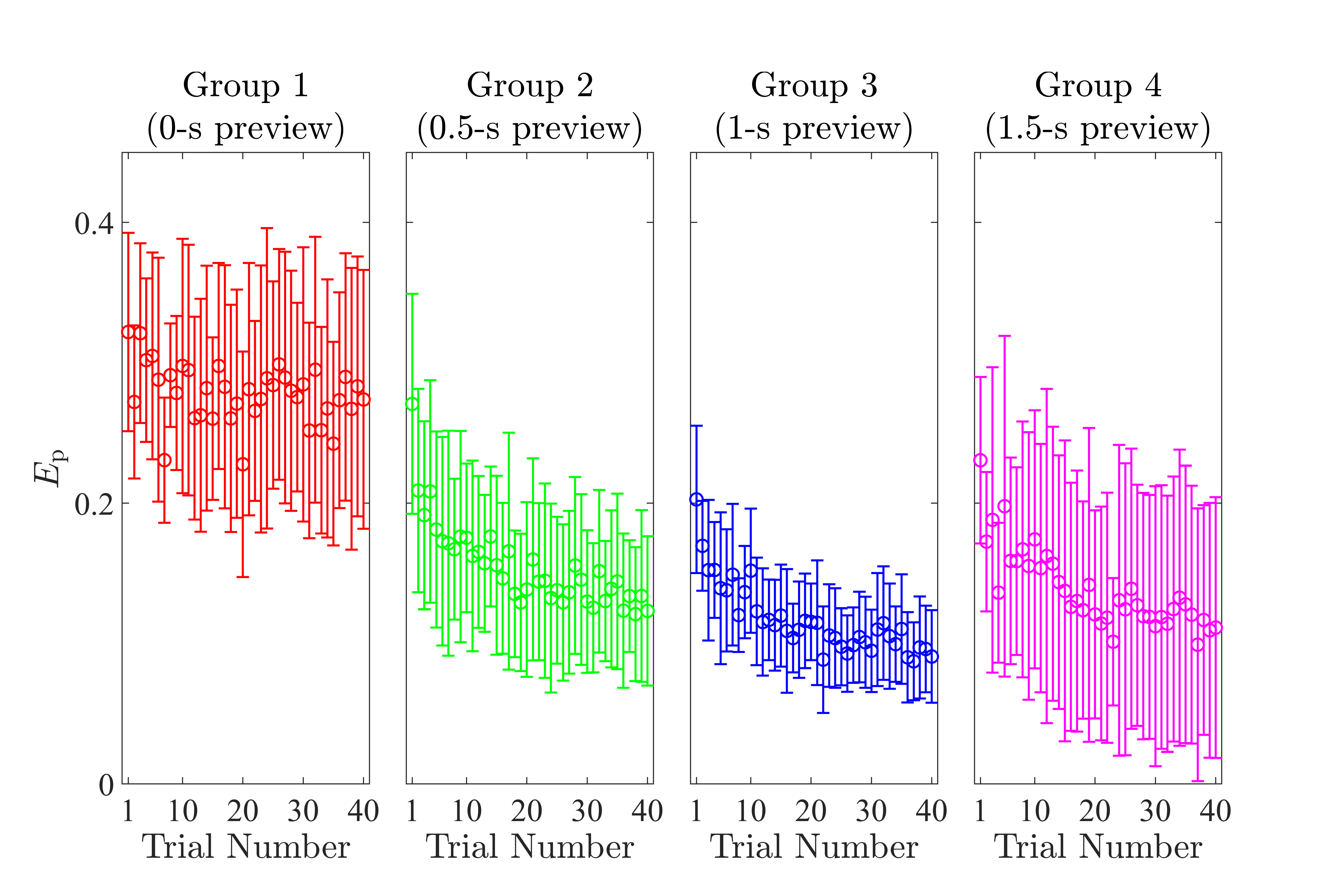}}
	\caption{
		Mean and standard deviation of $E_{\rm p}$ on each trial for each group.
		The $\circ$ is the mean, and the lines indicate the standard deviation.
		}\label{fig:Ep}
\end{figure}

Figures~\ref{fig:Em} and~\ref{fig:Ep} show the mean and standard deviation of $E_{\rm m}$ and $E_{\rm p}$ for each group on each trial. 
These results are similar to the time-domain results shown in Fig.~\ref{fig:error}. 
However, for each group, mean $E_{\rm p}$ is generally greater than mean $E_{\rm m}$, which suggests that the command-following error is a result of error in phase more than error in magnitude.

\begin{table}[t!]
\footnotesize
	\centering
	\caption{Mean $E_{\rm m} \times 10^{-2}$ and change from first 5 to last 5 trials}
	\label{table:Em}
	\begin{tabular}
	{C{0.29in} | C{0.27in} C{0.27in} C{0.28in} C{0.28in} C{0.28in} C{0.28in} | C{0.34in}}
     & Trials &  Trials  & Trials & Trials & Trials & Trials & \\
    Group & 1--5 &  6--10  & 11--20 & 21--30 & 31--35 & 36--40 & Change \\
		\hline
		1       & $12.6$ & $12.0$ & $11.2$ & $11.2$ & $10.5$ & $10.9$ & $-1.7$\\
		2       & $11.0$ & $10.0$ & $10.0$  & $8.8$  & $8.1$  & $7.9$  & $-3.0$\\
		3       & $10.1$ & $9.3$  & $7.9$  & $7.1$  & $7.2$  & $7.1$  & $-2.9$\\
		4       & $10.2$ & $9.0$  & $8.1$  & $7.5$  & $7.5$  & $7.1$  & $-3.1$\\
	\end{tabular}
\end{table}
\begin{table}[t!]
\footnotesize
	\centering
	\caption{Mean $E_{\rm p} \times 10^{-2}$ and change from first 5 to last 5 trials}
	\label{table:Ep}
	\begin{tabular}
	{C{0.29in} | C{0.27in} C{0.27in} C{0.28in} C{0.28in} C{0.28in} C{0.28in} | C{0.34in}}
     & Trials &  Trials  & Trials & Trials & Trials & Trials & \\
    Group & 1--5 &  6--10  & 11--20 & 21--30 & 31--35 & 36--40 & Change \\
		\hline
		1       & $30.7$ & $27.9$ & $27.0$ & $28.3$ & $26.3$ & $27.8$ & $-2.9$\\
		2       & $20.6$ & $17.3$ & $15.4$ & $14.2$ & $13.8$ & $12.7$ & $-7.9$\\
		3       & $16.1$ & $13.9$ & $11.4$ & $10.0$ & $10.8$ & $9.2$  & $-6.9$\\
		4       & $18.4$ & $16.3$ & $14.0$ & $12.1$ & $12.4$ & $11.2$ & $-7.3$\\
	\end{tabular}
\end{table}

Tables~\ref{table:Em} and~\ref{table:Ep} show mean $E_{\rm m}$ and mean $E_{\rm p}$ for each group on different sets of trials. 
For groups~2--4, mean $E_{\rm p}$ decreases more (in absolute and percent) than mean $E_{\rm m}$. 
A paired $t$-test of change in $E_{\rm m}$ and change in $E_{\rm p}$ from the first $5$ trials to the last $5$ trials yields $p<0.05$ for groups~2--4.
This result suggests that the improvement in $\|e\|$ for groups~2--4 is attributed more to improvement in matching the phase of the reference than improvement in matching its magnitude.

For group~1, mean $E_{\rm m}$ and mean $E_{\rm p}$ also decrease over the trials although the percent decrease in mean $E_{\rm m}$ is slightly greater than that of mean $E_{\rm p}$. 
We also note that mean $E_{\rm p}$ for group~1 is significantly greater than that for the other groups. 
Thus, group~1's mean $\|e\|$ is greater than that of the other groups primarily because of error in phase as opposed to error in magnitude.

\section{Identified Models of Control Strategies}

For each trial of each subject, we use the SSID algorithm in Section~\ref{sec:SSID} to obtain the best fit feedforward $z^{-\tau_{\rm ff}} G_{\rm ff}$ and feedback $z^{-\tau_{\rm fb}} G_{\rm fb}$ controllers. 
Appendix~\ref{app:D} presents a validation analysis of the SSID results.


\subsection{Feedforward Control}

\begin{figure}[b]
\center{\includegraphics[width=0.48\textwidth,clip=true,trim= 0.2in 0.1in 0.75in 0.48in]
	{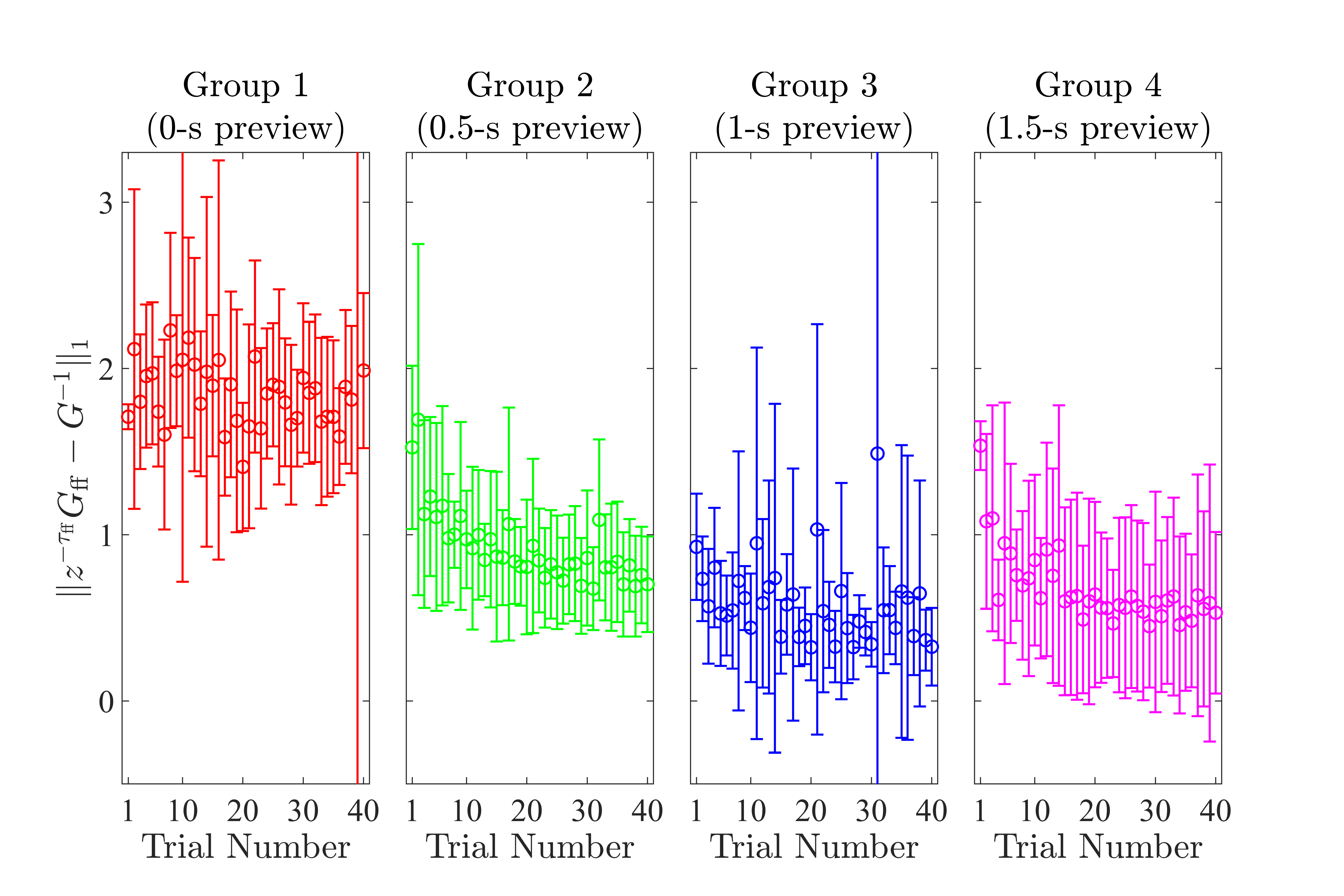}}
	\caption{
		Mean and standard deviation of $\|z^{-\tau_{\rm ff}}G_{\rm ff} - G^{-1}\|_1$ on each trial. 
		The $\circ$ is the mean, and the lines indicate the standard deviation.
	}
	\label{fig:Gff}
\end{figure}

For each identified feedforward controller, we define
\begin{align*}
\| z^{-\tau_{\rm ff}}G_{\rm ff} - G^{-1}  \|_{1}
&\triangleq \frac{1}{\pi} \int_{0}^{\pi} \left | e^{-\jmath \omega T_\rms \tau_{\rm ff}}G_{\rm ff}(e^{\jmath \omega T_\rms}) \right .\\ 
& \qquad \left . -G^{-1}(e^{\jmath \omega T_\rms}) \right | \, \rmd \omega,
\end{align*}
which is the frequency-averaged magnitude of the difference between the identified $z^{-\tau_{\rm ff}}G_{\rm ff}$ and the inverse dynamics $G^{-1}$ over the $0$-to-$0.5$ Hz range.
Figure~\ref{fig:Gff} shows the trial-by-trial mean and standard deviation of ${\| z^{-\tau_{\rm ff}}G_{\rm ff} - G^{-1}  \|_{1}}$ for each group, and Table~\ref{table:Gff} provides mean $\| z^{-\tau_{\rm ff}}G_{\rm ff} - G^{-1} \|_{1}$ for each group on different sets of trials.

The mean ${\| z^{-\tau_{\rm ff}}G_{\rm ff} - G^{-1}  \|_{1}}$ for groups~2--4 decrease over the trials.
Table~\ref{table:Gff} shows that mean $\| z^{-\tau_{\rm ff}}G_{\rm ff} - G^{-1}  \|_1$ decreases by $43\%$, $33\%$, and $46\%$ from the first 5 trials to the last 5 trials for groups~2--4, respectively.
In contrast, mean $\| z^{-\tau_{\rm ff}}G_{\rm ff} - G^{-1} \|_1$ for group~1 does not decrease over the trials and is significantly larger than that of the other groups over the last 30 trials.

\begin{table}[t]
\footnotesize
	\centering
	\caption{Mean $\| z^{-\tau_{\rm ff}}G_{\rm ff} - G^{-1} \|_1 $}
	\label{table:Gff}
	\begin{tabular}
    {C{0.29in} | C{0.26in} C{0.26in} C{0.28in} C{0.28in} C{0.28in} C{0.28in} }
     & Trials &  Trials  & Trials & Trials & Trials & Trials \\
    Group & 1--5 &  6--10  & 11--20 & 21--30 & 31--35 & 36--40 \\
		\hline
		1       & $1.92$ & $1.91$ & $1.85$ & $1.81$ & $1.77$ & $2.17$ \\
		2       & $1.28$ & $1.05$ & $0.90$ & $0.81$ & $0.83$ & $0.73$ \\
		3       & $0.70$ & $0.57$ & $0.57$ & $0.50$ & $0.75$ & $0.47$ \\
		4       & $1.04$ & $0.77$ & $0.68$ & $0.55$ & $0.55$ & $0.56$ \\
	\end{tabular}
\end{table}

\begin{figure}[t]
	\center{\includegraphics[width=0.48\textwidth,clip=true,trim= 0.2in 0.05in 0.88in 0.46in]
		{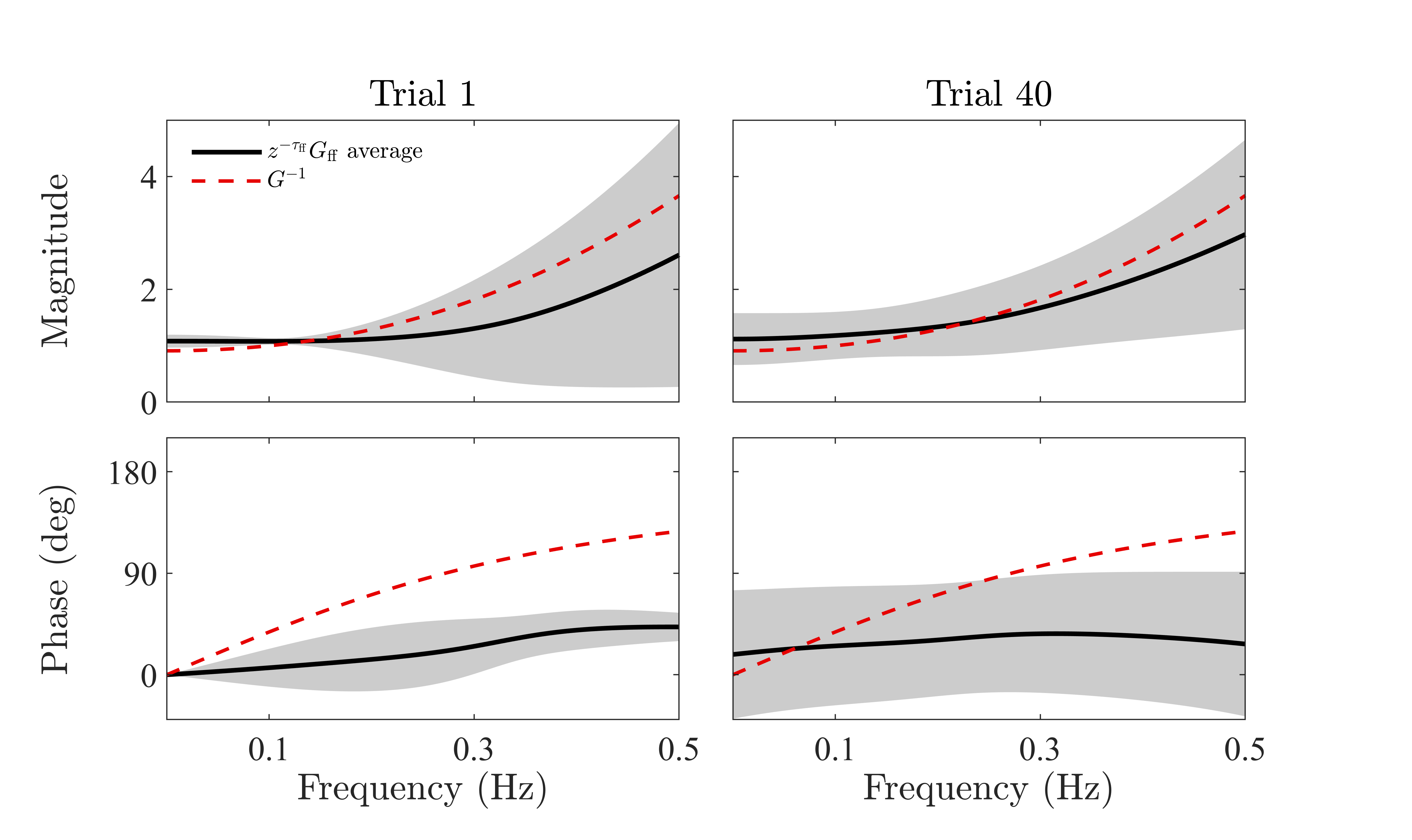}}
	\caption{
		Average identified feedforward controller for group~1 on trials 1 and 40.
		The shaded region shows the standard deviation.
	}
	\label{fig:group1bode}
\end{figure}

\begin{figure}[t!]
	\center{
 \includegraphics[width=0.48\textwidth,clip=true,trim= 0.2in 0.05in 0.88in 0.46in]
		{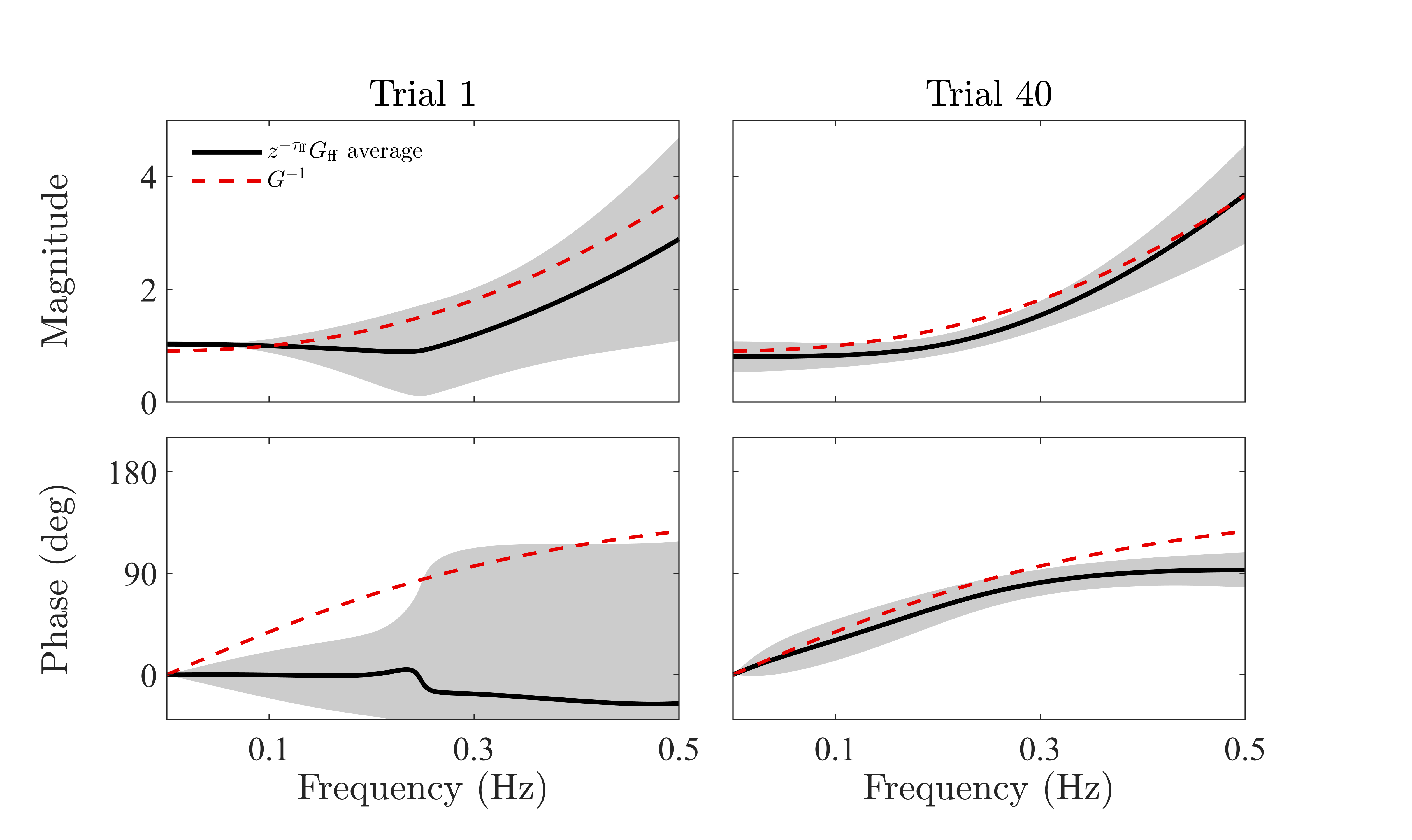}}
	\caption{
		Average identified feedforward controller for group~2 on trials 1 and 40.
		The shaded region shows the standard deviation.
	}
	\label{fig:group2bode}
\end{figure}

\begin{figure}[t!]
	\center{
 \includegraphics[width=0.48\textwidth,clip=true,trim= 0.2in 0.05in 0.88in 0.46in]
		{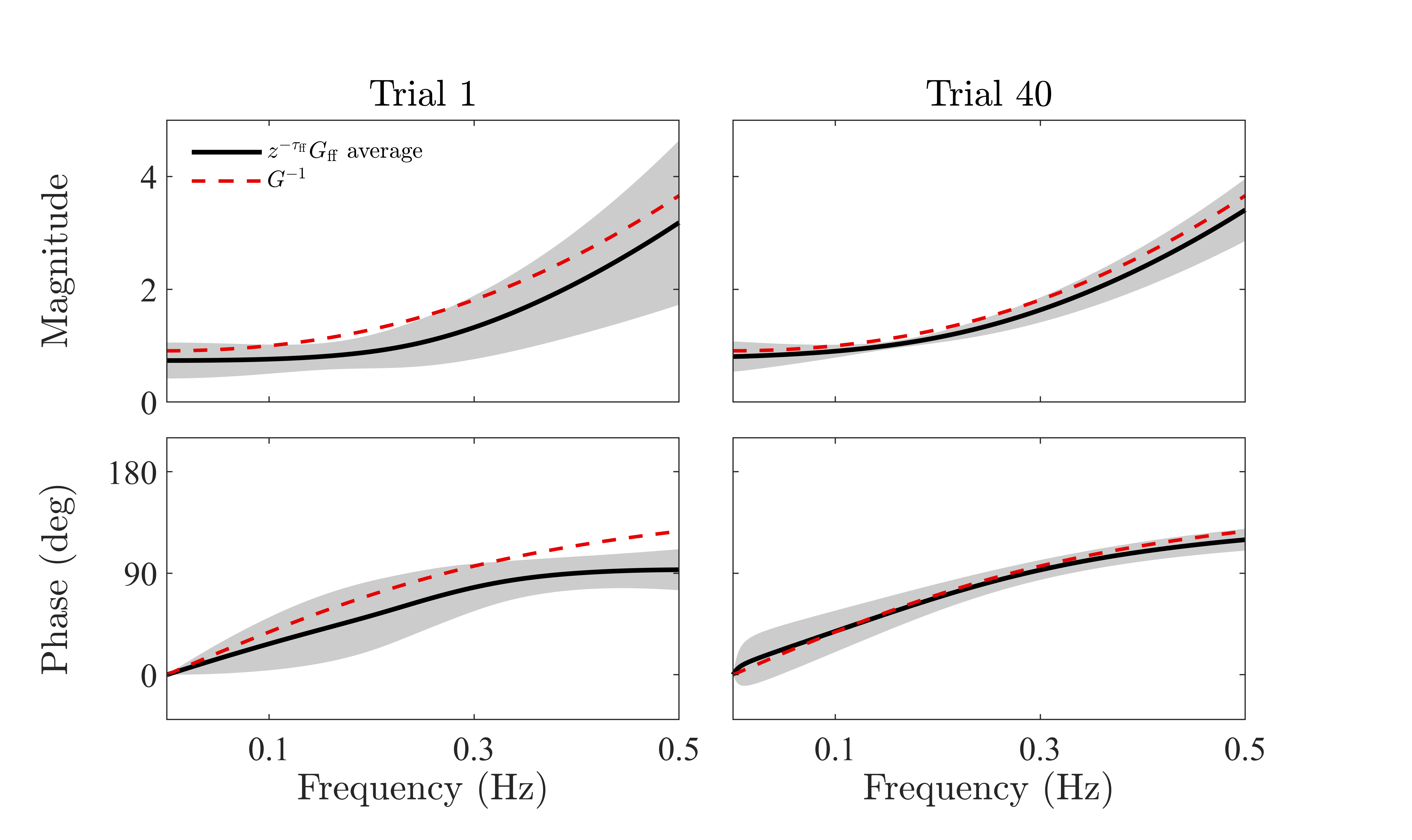}}
	\caption{
		Average identified feedforward controller for group~3 on trials 1 and 40.
		The shaded region shows the standard deviation.
	}
	\label{fig:group3bode}
\end{figure}

\begin{figure}[t!]
	\center{
 \includegraphics[width=0.48\textwidth,clip=true,trim= 0.2in 0.05in 0.88in 0.46in]
		{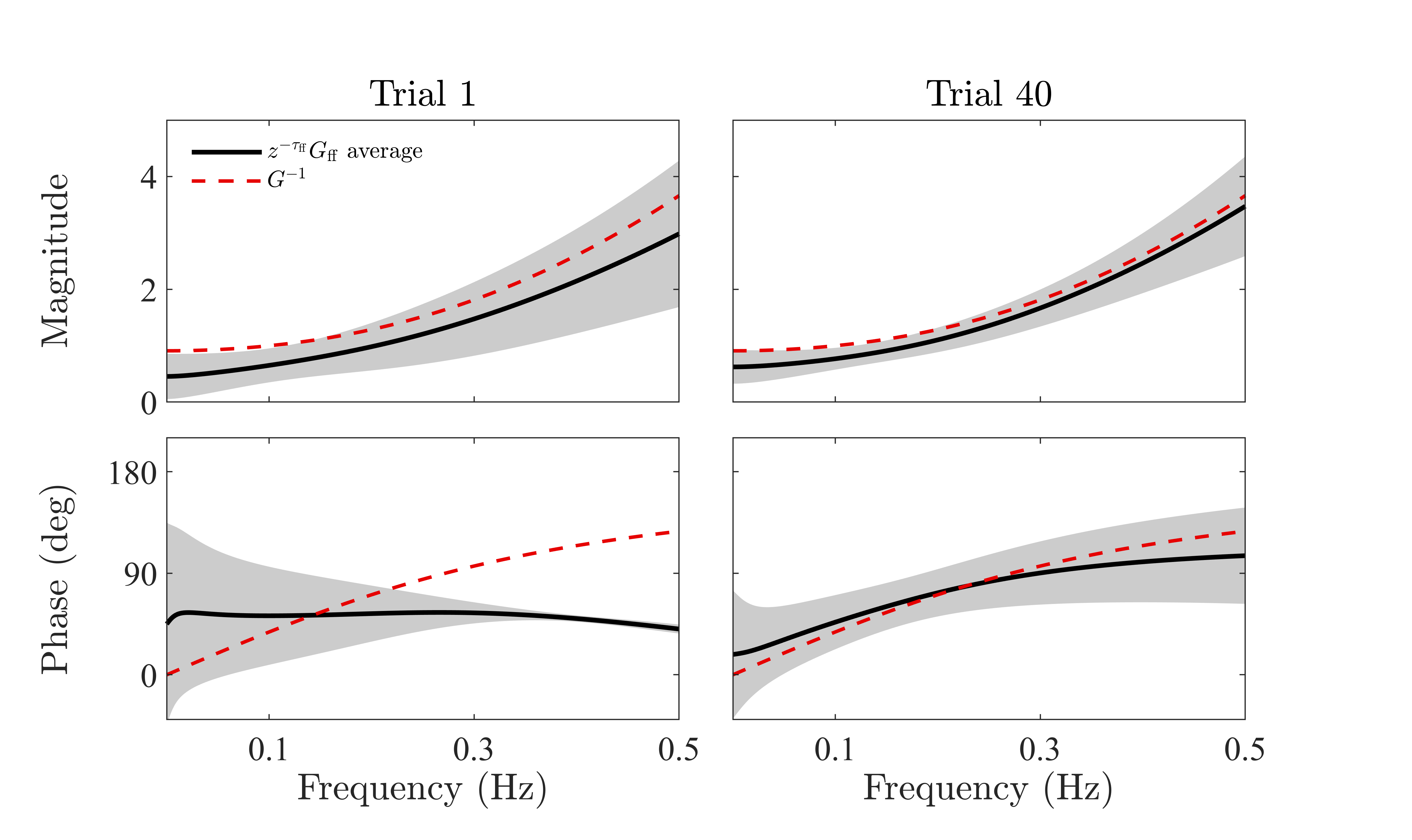}}
	\caption{
		Average identified feedforward controller for group~4 on trials 1 and 40.
		The shaded region shows the standard deviation.
	}
	\label{fig:group4bode}
\end{figure}

Figures~\ref{fig:group1bode}--\ref{fig:group4bode} are the Bode plots of the average identified feedforward controller $z^{-\tau_{\rm ff}}G_{\rm ff}$ over all 11 subjects on the first and last trials for groups~1--4, respectively.
For groups~2--4, the average identified $z^{-\tau_{\rm ff}}G_{\rm ff}$ on trial~40 approximates $G^{-1}$ better than on trial~1.
These results indicate that the subjects in groups 2--4 learn to use an approximation of the inverse system dynamics $G^{-1}$ in feedforward. 
Hence, the results for groups~2--4 demonstrate that using the approximate inverse dynamics in feedforward is a primary control strategy not only with predictable commands \cite{zhang2018a, zhang2022a, seyyedmousavi2020b, seyyedmousavi2020c, Koushkbaghi_JFI_Nonlinear} but also with unpredictable commands if preview of the command is provided. 
In addition, the results for group~1 suggest that if the reference command is unpredictable, then preview is necessary to approximate inverse dynamics in feedforward.
In other words, the subjects in group~1 do not learn to approximate the inverse dynamics in feedforward. 

\subsection{Feedback Control} 
\label{sec:Feedback}

\begin{figure}[t]
	\center{\includegraphics[width=0.48\textwidth,clip=true,trim= 0.2in 0.1in 0.75in 0.48in]
	{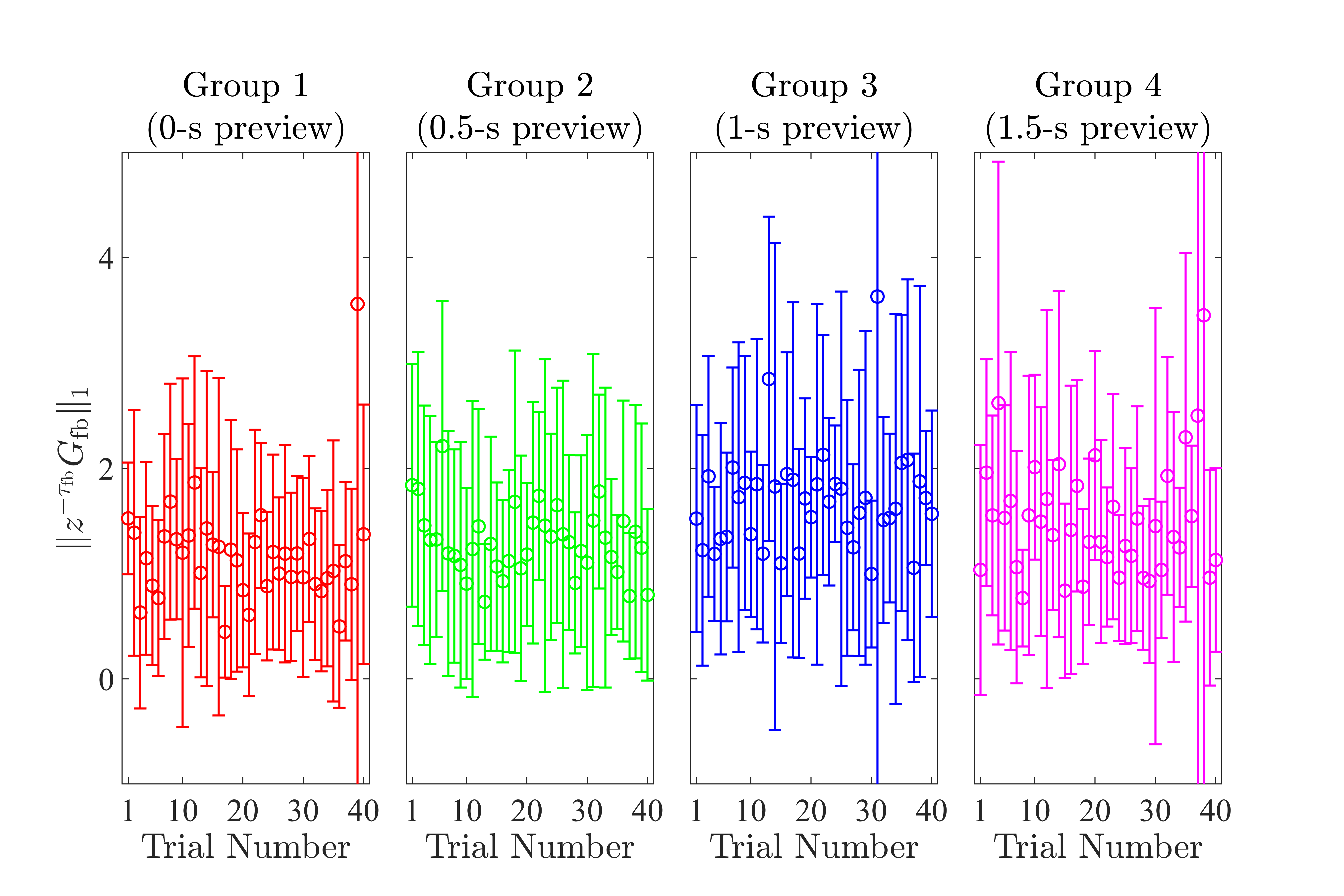}}
	\caption{
		Mean and standard deviation of $\| z^{-\tau_{\rm fb}}G_{\rm fb} \|_1$ on each trial. 
		The $\circ$ is the mean, and the lines indicate the standard deviation.  
	}
	\label{fig:Gfb}
\end{figure}

For each identified feedback controller, we define 
\begin{equation*}
\| z^{-\tau_{\rm fb}}G_{\rm fb} \|_1 \triangleq \frac{1}{\pi} \int_{0}^{\pi}  \left | e^{-\jmath \omega T_{\rm s} \tau_{\rm fb}} G_{\rm fb}(e^{\jmath \omega T_{\rm s}}) \right | \, \rmd \omega,
\end{equation*}
which is the frequency-averaged magnitude of $z^{-\tau_{\rm fb}}G_{\rm fb}$ over the $0$-to-$0.5$~Hz range.
Figure~\ref{fig:Gfb} shows the trial-by-trial mean and standard deviation of $\| z^{-\tau_{\rm fb}}G_{\rm fb} \|_1$ for each group, and Table~\ref{table:Gfb} shows mean $\| z^{-\tau_{\rm fb}}G_{\rm fb} \|_1$ for each group on different sets of trials.

The mean $\| z^{-\tau_{\rm fb}}G_{\rm fb} \|_1$ for group~1 is larger on the last 5 trials than on the early trials. 
In particular, mean $\| z^{-\tau_{\rm fb}}G_{\rm fb} \|_1$ increases $46$\% from the first 5 to the last 5 trials, which suggests that subjects in group~1 may learn to increase feedback gain over the trials.
This observation may partly explain the mechanism that group~1 subjects use to improve performance (mean $\| e \|$ decreases by $16$\% from first 5 to last 5 trials) even though they were unable to learn to use the approximate inverses dynamics in feedforward.


The average identified feedback time delays $T_{\rm fb}$ for groups~1--4 over all 40~trials are $288$~ms, $294$~ms, $266$~ms, and $275$~ms, respectively. 
There is no apparent trend in the mean or standard deviation of $T_{\rm fb}$ over the trials.
These results on feedback time delay are consistent with \cite{zhang2018a, seyyedmousavi2020b}.

\begin{table}[t]
\footnotesize
	\centering
	\caption{Mean $\|z^{-\tau_{\rm fb}}G_{\rm fb}\|_1$}
	\label{table:Gfb}
	\begin{tabular}
	{C{0.29in} | C{0.26in} C{0.26in} C{0.28in} C{0.28in} C{0.28in} C{0.28in} }
     & Trials &  Trials  & Trials & Trials & Trials & Trials \\
    Group & 1--5 &  6--10  & 11--20 & 21--30 & 31--35 & 36--40  \\
		\hline
		1       & $1.04$ & $1.24$ & $1.18$ & $1.08$ & $1.00$ & $1.52$ \\
		2      & $1.49$ & $1.33$ & $1.16$ & $1.35$ & $1.33$ & $1.14$ \\
		3       & $1.46$ & $1.67$ & $1.69$ & $1.63$ & $2.09$ & $1.66$ \\
		4     & $1.74$ & $1.35$ & $1.50$ & $1.23$ & $1.56$ & $1.93$ \\
	\end{tabular}
\end{table}

\section{Effect of Preview on the Approximation of $G^{-1}$ in Feedforward}

Groups~2--4 learn to approximate $G^{-1}$ in feedforward; however, Fig.~\ref{fig:Gff} and Table~\ref{table:Gff} show that these groups do not learn the approximation equally well. 
For the last 5 trials, mean ${\| z^{-\tau_{\rm ff}}G_{\rm ff} - G^{-1}  \|_{1}}$ is smallest for group~3 followed in order by groups~4, 2, and~1.
Table~\ref{table:error} shows that mean $\| e \|$ for the last 5 trials is in the same order from smallest to largest (group~3, 4, 2, 1). 
Thus, for the last 5 trials, the group with 1-s preview (i.e., group~3) has the best approximation of $G^{-1}$ in feedforward and the best time-domain performance.
Group~4 with 1.5-s preview is second best; group~2 with 0.5-s preview is third best; and group~1 with no preview is worst. 
This observation suggests that group~2's 0.5-s preview may be too little to allow the subject's to learn as accurate an approximation of $G^{-1}$ as the one learned by group~3. 
It is also possible that group~4 may have too much preview in comparison to group~3. 
We note that the same smallest-to-largest order (group~3, 4, 2, 1) holds for mean $\| e \|$ and mean ${\| z^{-\tau_{\rm ff}}G_{\rm ff} - G^{-1}  \|_{1}}$ on all sets of trials in Tables~\ref{table:error} and~\ref{table:Gff} except for trials 31--35 where group~4 has the smallest mean ${\| z^{-\tau_{\rm ff}}G_{\rm ff} - G^{-1}  \|_{1}}$.

A key component of learning to approximate $G^{-1}$ in feedforward is learning to use the correct amount of phase lead in feedforward \cite{seyyedmousavi2020b}.
Figures~\ref{fig:group2bode}--\ref{fig:group4bode} show that for groups~2--4, the average identified feedforward controller has significant phase lag relative to $G^{-1}$  on trial 1, and this phase lag is substantially reduced (or eliminated) by trial 40.
In contrast, the average identified feedforward controller for group~1 (Fig.~\ref{fig:group1bode}) has significant phase lag relative to $G^{-1}$ not only on the first trial but also on the last trial.

\begin{figure}[t!]
	\center{\includegraphics[width=0.48\textwidth,clip=true,trim= 0.14in 0.1in 0.75in 0.48in]
	{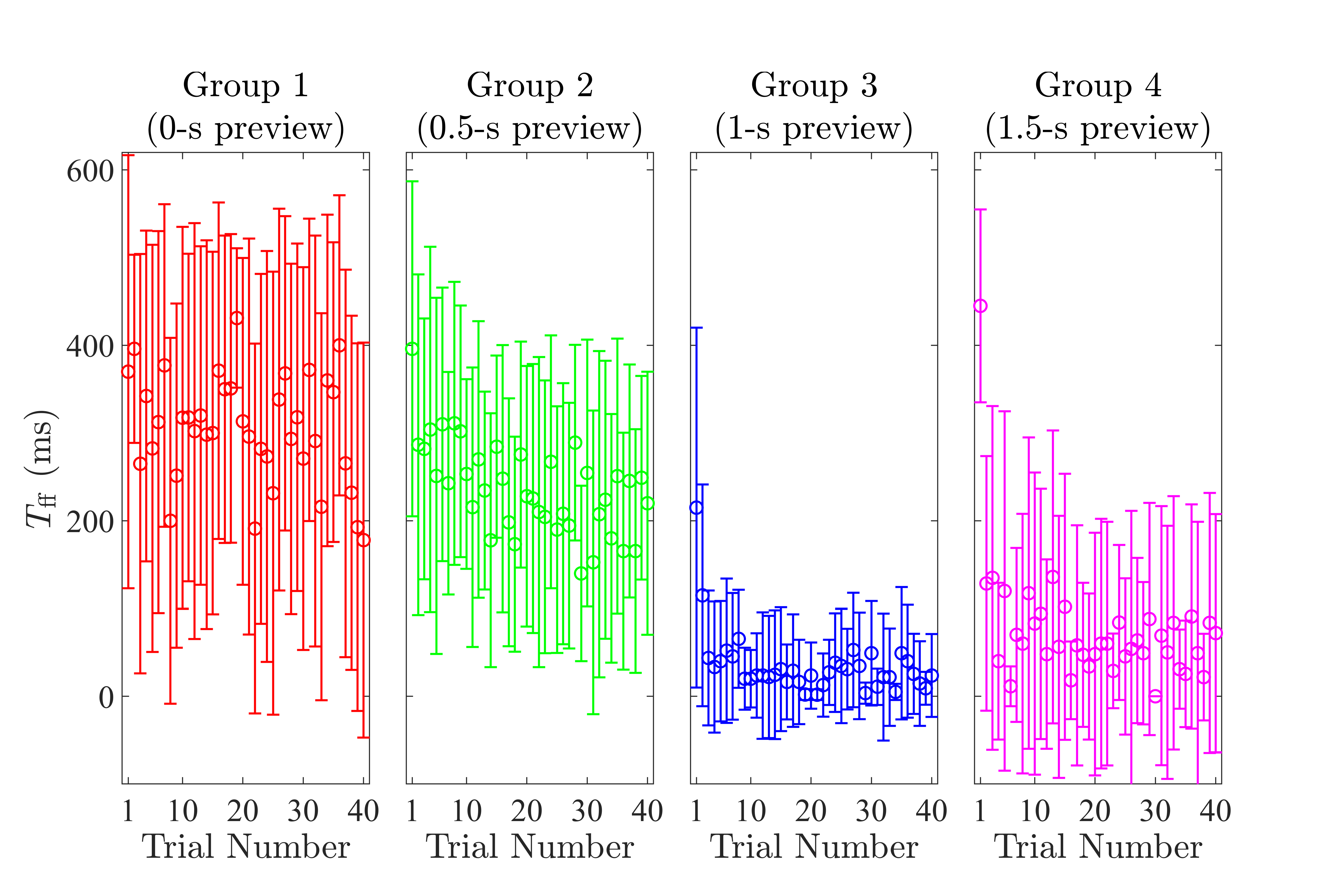}}
	\caption{
		Mean and standard deviation of $T_{\rm ff}$ on each trial.
		The $\circ$ is the mean, and the lines indicate the standard deviation. 
	}
	\label{fig:Tff}
\end{figure}

Time delay is one important characteristic that can cause phase lag in the feedforward control $z^{-\tau_{\rm ff}} G_{\rm ff}$. 
Figure~\ref{fig:Tff} shows the trial-by-trial mean and standard deviation of the identified feedforward time delay $T_{\rm ff}$ for each group, and Table~\ref{table:Tff} shows mean $T_{\rm ff}$ for each group on different sets of trials.

\begin{table}[t]
\footnotesize
	\centering
	\caption{Mean $T_{\rm ff}$}
	\label{table:Tff}
	\begin{tabular}
{C{0.29in} | C{0.26in} C{0.26in} C{0.28in} C{0.28in} C{0.28in} C{0.28in} }
     & Trials &  Trials  & Trials & Trials & Trials & Trials \\
    Group & 1--5 &  6--10  & 11--20 & 21--30 & 31--35 & 36--40  \\
		\hline
		1       & $321$ & $296$ & $335$ & $289$ & $316$ & $253$ \\
		2       & $296$ & $287$ & $231$ & $219$ & $202$ & $209$ \\
		3       & $83$  & $41$  & $21$  & $29$  & $23$  & $23$  \\
		4       & $155$ & $69$  & $63$  & $54$  & $52$  & $63$  \\
		
	\end{tabular}
\end{table}

For groups~2--4, mean $T_{\rm ff}$ decreases over the trials; in particular, mean $T_{\rm ff}$ for groups~2--4 decreases by $29$\%, $72$\%, and $59$\%, respectively, from the first 5 trials to the last 5 trials.
In contrast, mean $T_{\rm ff}$ for group~1 is comparatively large and does not have a consistent trend over trials (although it is smallest over the last five trials).

For the last 5 trials, mean $T_{\rm ff}$ is smallest for group~3 followed in order by groups~4, 2, and~1. 
This is the same order observed for mean ${\| z^{-\tau_{\rm ff}}G_{\rm ff} - G^{-1}  \|_{1}}$.
In fact, the same smallest-to-largest order (group~3, 4, 2, 1) holds for mean $T_{\rm ff}$ on all sets of trials in Table~\ref{table:Tff}.

The results in Figure~\ref{fig:Tff} and Table~\ref{table:Tff} suggest that subjects use preview to compensate for sensory time delay and reduce their effective feedforward time delay $T_{\rm ff}$.
Furthermore, this ability to reduce time delay in feedforward is a critical to being able to implement the necessary phase lead to approximate $G^{-1}$ in feedforward.

\begin{figure}[t]
\center{\includegraphics[width=0.48\textwidth,clip=true,trim= 0.1in 0.05in 0.85in 0.46in]
		{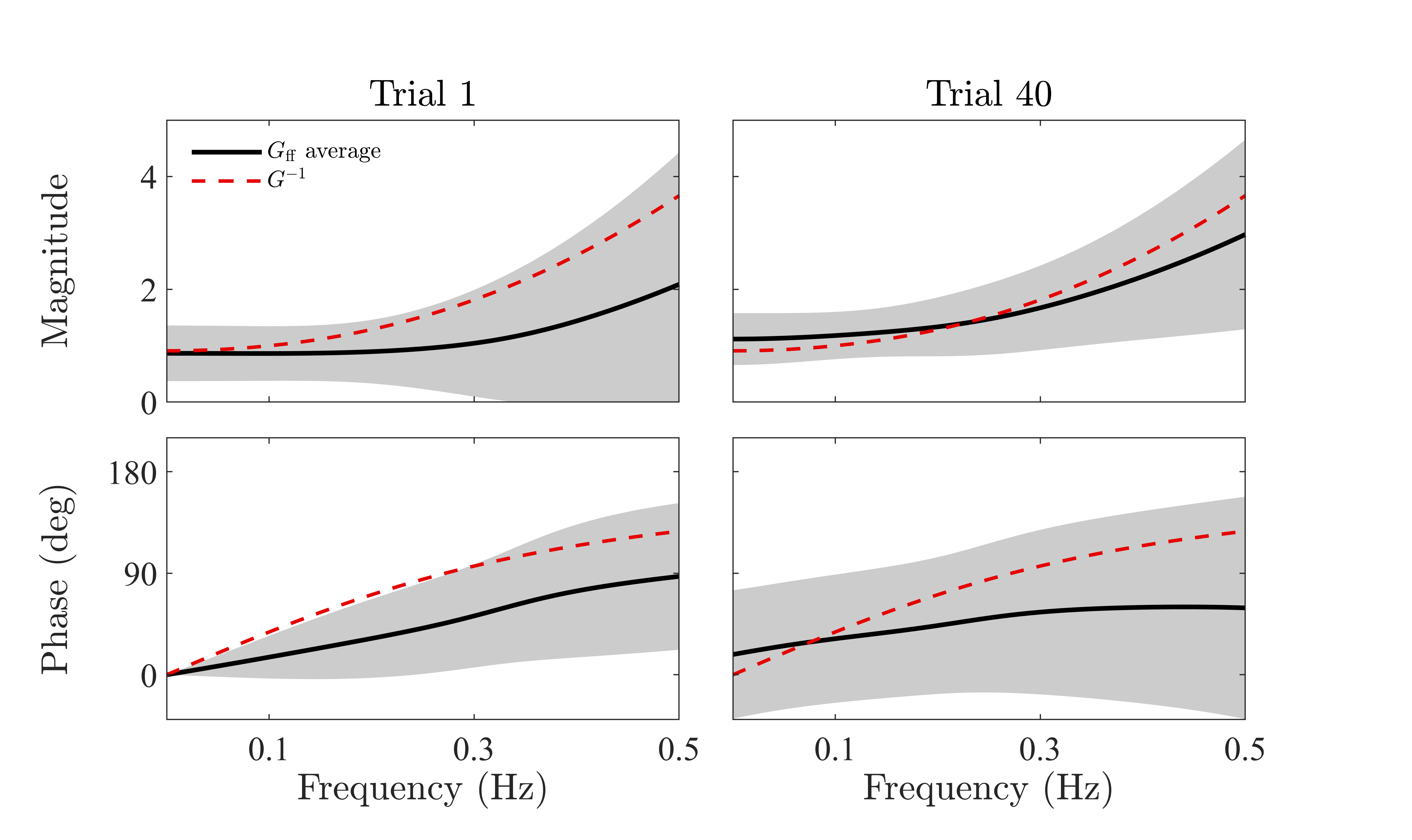}}
	\caption{
		Average identified feedforward transfer function for group~1 on trials 1 and 40.
		The shaded region shows the standard deviation.
	}
	\label{fig:group1bode_nodelay}
\end{figure}

In contrast, group~1 cannot compensate for sensory time delay because the reference is unpredictable and they do not have preview. 
Thus, these subjects are limited in their ability to reduce their effective feedforward time delay $T_{\rm ff}$, which, in turn, limits their ability to implement the phase lead needed to approximate $G^{-1}$ in feedforward.
To elucidate this point further, consider Figure~\ref{fig:group1bode_nodelay}, which is the Bode plot of the average identified $G_{\rm ff}$ over all 11~subjects on the first and last trial for group~1.
Note that if the feedforward delay is zero (i.e., $\tau_{\rm ff}=0$), then $G_{\rm ff}=z^{-\tau_{\rm ff}} G_{\rm ff}$. 
Thus, we can interpret $G_{\rm ff}$ as the feedforward controller that the subjects would achieve if they could eliminate their time delay in feedforward.
Comparing Figs.~\ref{fig:group1bode} and~\ref{fig:group1bode_nodelay} shows that the phase of $G^{-1}$ is better approximated by the phase of $G_{\rm ff}$ than by the phase of $z^{-\tau_{\rm ff}} G_{\rm ff}$. 
This insight suggests that group~1 may be attempting to approximating $G^{-1}$ in feedforward; however, the significant feedforward time delay (see Fig.~\ref{fig:Tff}) prevents an accurate approximation of the phase of $G^{-1}$.
Even if the subjects in group~1 could eliminate their time delay in feedforward, then the average identified feedforward behavior (Fig.~\ref{fig:group1bode_nodelay}) would still have phase lag relative to $G^{-1}$.
This suggests that the absence of preview not only limits the ability to reduce effective feedforward time delay but also imposes limits on the amount of phase lead that can be implemented in feedforward.

Next, we define metrics that quantify how much of the difference between $z^{-\tau_{\rm ff}}G_{\rm ff}$ and $G^{-1}$ is attributed to difference in magnitude versus difference in phase. 
For each trial, the \textit{frequency-averaged error in the magnitude of $z^{-\tau_{\rm ff}} G_{\rm ff}G$ relative to unity} is
\begin{equation*}
M_{\rm e}(z^{-\tau_{\rm ff}} G_{\rm ff}) 
\triangleq \frac{1}{\pi} \int_{0}^{\pi} \bigg | \left | e^{-\jmath \omega T_\rms \tau_{\rm ff}} G_{\rm ff}(e^{\jmath \omega T_\rms}) G(e^{\jmath \omega T_\rms}) \right | - 1 \bigg | \, \rmd \omega,
\end{equation*}
which is the frequency-averaged magnitude of the difference between $z^{-\tau_{\rm ff}} G_{\rm ff}G$ and $1$ assuming that the phase of $z^{-\tau_{\rm ff}} G_{\rm ff}G$ is equal to the phase of $1$. 
For each trial, the \textit{frequency-averaged error in the phase of $z^{-\tau_{\rm ff}} G_{\rm ff}G$ relative to unity} is
\begin{equation*}
P_{\rm e}(z^{-\tau_{\rm ff}} G_{\rm ff}) 
\triangleq \frac{1}{\pi} \int_{0}^{\pi}  \bigg | e^{j\angle [ e^{- \jmath \omega T_\rms \tau_{\rm ff}} G_{\rm ff}(e^{\jmath \omega T_\rms}) G(e^{\jmath \omega T_\rms })]} -1 \bigg| \, \rmd \omega,
\end{equation*} 
which is the frequency-averaged magnitude of the difference between $z^{-\tau_{\rm ff}} G_{\rm ff}G$ and $1$ assuming that the magnitude of $z^{-\tau_{\rm ff}} G_{\rm ff}G$ is equal to the magnitude of $1$. 
We compare $M_\rme$ and $P_\rme$ to determine if the difference between $z^{-\tau_{\rm ff}} G_{\rm ff} G$ and $1$ is due more to error in magnitude or error in phase.

\begin{figure}[t]
\center{\includegraphics[width=0.48\textwidth,clip=true,trim= 0.2in 0.1in 0.75in 0.48in] 
		{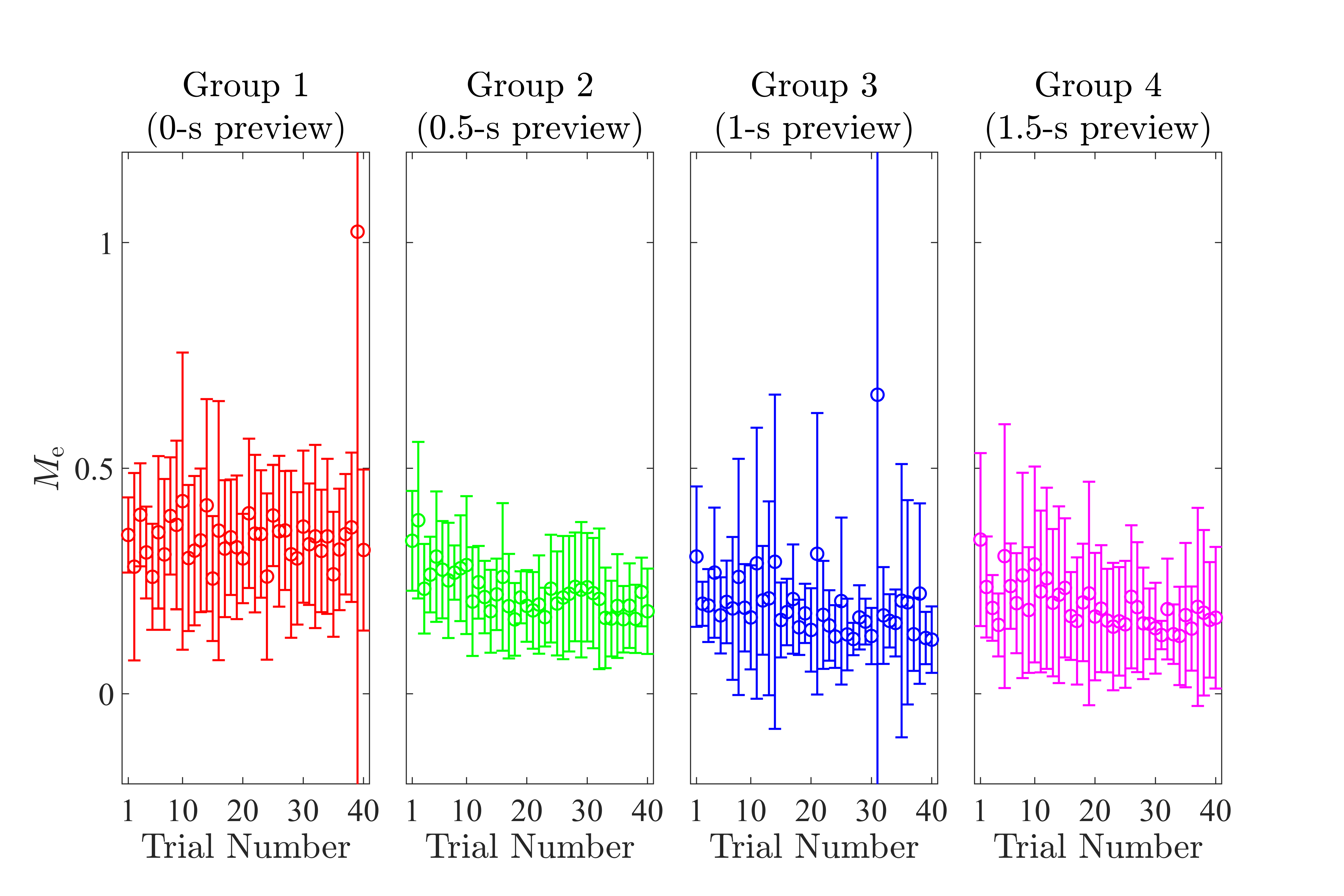}}
	\vspace{-1ex}
	\caption{
		Mean and standard deviation of $M_{\rm e}$ on each trial. 
		The $\circ$ is the mean, and the lines indicate standard deviation.
	}
	\label{fig:Me}
\end{figure}

\begin{figure}[t]
	\center{\includegraphics[width=0.48\textwidth,clip=true,trim= 0.2in 0.1in 0.75in 0.48in] 
		{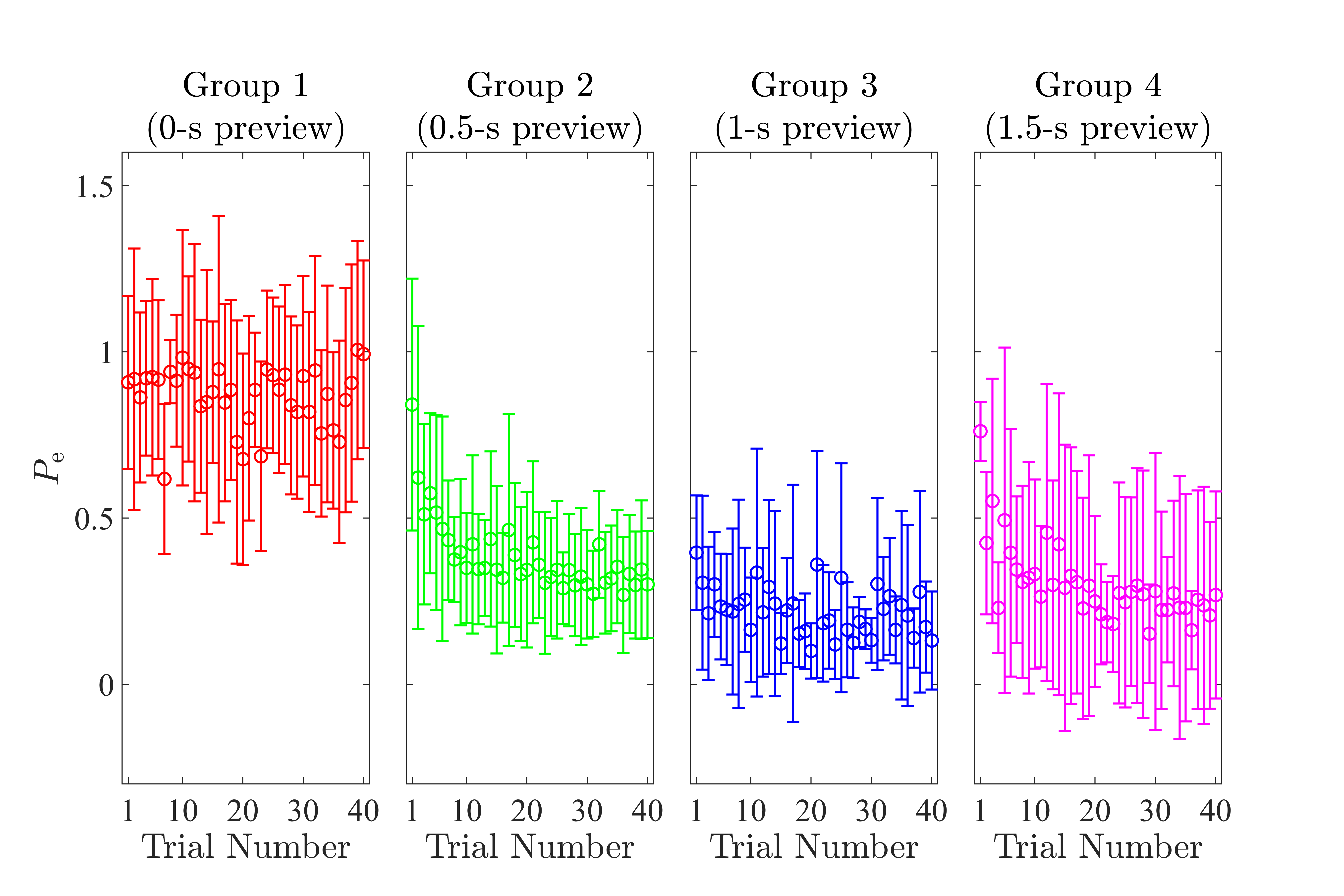}}
	\vspace{-1ex}
	\caption{
		Mean and standard deviation of $P_{\rm e}$ on each trial. 
		The $\circ$ is the mean, and the lines indicate standard deviation.
	}
	\label{fig:Pe}
\end{figure}

Figures~\ref{fig:Me} and~\ref{fig:Pe} show the trial-by-trail mean and standard deviation of $M_{\rm e}$ and $P_{\rm e}$ for each group, and Tables~\ref{table:Me} and~\ref{table:Pe} show the mean $M_{\rm e}$ and mean $P_{\rm e}$ for each group on different sets of trials.
For groups~2--4, mean $M_{\rm e}$ and mean $P_{\rm e}$ decrease over the trials.
In contrast, for group~1, mean $P_{\rm e}$ does not change significantly over the trials and mean $M_{\rm e}$ increases on the last 5 trials.
Notably, Tables~\ref{table:Me} and~\ref{table:Pe} show that for groups 2--4, mean $P_{\rm e}$ decreases more (in absolute and percent) than the mean $M_{\rm e}$. 
Specifically, mean $M_{\rm e}$ decreases by $34\%$, $27\%$, and $29\%$ from the first 5 trials to the last 5 trials for groups~2--4, whereas mean $P_{\rm e}$ decreases by $47\%$, $32\%$, and $52\%$ in the same order. 
Thus, matching the magnitude of $G^{-1}$ in feedforward is attributed more to improvement in matching the phase of $G^{-1}$ than improvement in matching the magnitude of $G^{-1}$. 
This analysis supports the conclusion that learning the phase lead of $G^{-1}$ is a critical aspect of learning to approximate $G^{-1}$ in feedforward.

\begin{table}[t]
\footnotesize
\centering
	\caption{Mean $M_{\rm e}\times 10^{-2}$ and change from the first 5 to last 5 trials}
	\label{table:Me}
	\begin{tabular}
		{C{0.29in} | C{0.26in} C{0.26in} C{0.28in} C{0.28in} C{0.28in} C{0.28in} | C{0.34in}}
     & Trials &  Trials  & Trials & Trials & Trials & Trials & \\
    Group & 1--5 &  6--10  & 11--20 & 21--30 & 31--35 & 36--40 & Change \\
		\hline
		1       & $32$ & $37$ & $33$ & $35$ & $32$ & $49$ & $17$\\
		2       & $29$ & $27$ & $21$ & $21$ & $19$ & $19$ & $-10$\\
		3       & $22$ & $20$ & $20$ & $17$ & $28$ & $16$ & $-6$\\
		4       & $24$ & $24$ & $21$ & $17$ & $15$ & $17$ & $-7$\\
	\end{tabular}
\end{table}
\begin{table}[t]
	\footnotesize
	\centering
	\caption{Mean $P_{\rm e}\times 10^{-2}$ and change from first 5 to last 5 trials}
	\label{table:Pe}
	\begin{tabular}
		{C{0.29in} | C{0.26in} C{0.26in} C{0.28in} C{0.28in} C{0.28in} C{0.28in} | C{0.34in}}
     & Trials &  Trials  & Trials & Trials & Trials & Trials & \\
    Group & 1--5 &  6--10  & 11--20 & 21--30 & 31--35 & 36--40 & Change \\
		\hline
		1       & $91$ & $88$ & $85$ & $86$ & $83$ & $90$ & $-1$\\
		2       & $59$ & $40$ & $37$ & $33$ & $33$ & $31$ & $-28$\\
		3       & $28$ & $22$ & $21$ & $20$ & $24$ & $19$ & $-9$\\
		4       & $48$ & $34$ & $31$ & $24$ & $24$ & $23$ & $-25$\\
	\end{tabular}
\end{table}

\section{Summary and Discussion}

This article presented new results on the impact of reference-command preview. 
The experimental results demonstrate that preview helps to improve command-following performance (Fig.~\ref{fig:error}, Table~\ref{table:error}). 
Specifically, no preview or too little preview can be detrimental to performance. 
The experimental results also suggest that too much preview could be detrimental to performance, which means that there may be an optimal preview time that depends on the dynamic system being controlled. 
However, the difference in command-following performance between the best-performing group (group~3 with 1-s preview) and the second-best-performing group (group~4 with 1.5-s preview) is not statistically significant. 
Thus, additional investigation is needed to determine if too much preview is detrimental. 

The frequency-domain analysis demonstrates that the improvement in command-following performance for the groups with preview is attributed more to improvement in matching the phase of the command than to improvement in matching its magnitude (Figs.~\ref{fig:Em},~\ref{fig:Ep} and Tables~\ref{table:Em},~\ref{table:Ep}). 
This analysis also shows that the group without preview performs worse than other groups primarily because the lack of preview prevents subjects from learning to match the phase of the command.

The SSID results demonstrate that the groups with preview improve command-following performance by learning to approximate the inverse dynamics $G^{-1}$ in feedforward (Fig.~\ref{fig:Gff}, Table~\ref{table:Gff}). 
Thus, feedforward-dynamic-inversion control (which is observed with predictable commands in \cite{zhang2018a, zhang2022a, seyyedmousavi2020b, seyyedmousavi2020c, Koushkbaghi_JFI_Nonlinear}) is also used with unpredictable commands if sufficiently long preview of the command is provided to the human. 
The SSID results for the group without preview suggest that if the command is unpredictable, then preview is necessary for implementing an approximation of $G^{-1}$ in feedforward.

The subjects with preview improve command-following performance by improving the accuracy of their approximation of $G^{-1}$ in feedforward, and the accuracy of that approximation is driven, in large part, by learning to implement the correct phase lead in feedforward (Figs.~\ref{fig:Me},~\ref{fig:Pe} and Tables~\ref{table:Me},~\ref{table:Pe}). 
Furthermore, implementing the correct phase lead is directly connected to the use of preview to compensate for sensory time delay in feedforward (Fig.~\ref{fig:Tff}, Table~\ref{table:Tff}).


\appendices
\gdef\thesection{\Alph{section}}

\section{Validation of SSID Results} \label{app:D}

For each trial, we simulate the identified closed-loop system, where the input to the simulation is $\{r_k\}_{k=1}^n$, and the output of the simulation is the validation data $\{ y_{{\rmv},k} \}_{k=1}^n$.
Specifically, we simulate
$\hat y_{\rmv}(z) = \tilde G_{yr}(z) \hat r(z)$,
where all initial conditions are zero, $\hat y_{\rmv}(z)$ is the $z$-transform of the validation data $y_{{\rmv},k}$, and $\tilde G_{yr}$ is the closed-loop transfer function \eqref{eq:tildeGyr} obtained from the identified $G_{\rm ff}$, $\tau_{\rm ff}$, $G_{\rm fb}$, and $\tau_{\rm fb}$.

For each trial, we compute the variance accounted for (VAF), which is a measure of the accuracy of the identified closed-loop transfer function and is given by
\begin{equation*}
	{\rm VAF} \triangleq  1-\frac{\sum_{k=n_{1}}^{n} |y_k - y_{{\rm v},k}|^2}{\sum_{k=n_{1}}^{n} |y_k|^2 },
\end{equation*}
where $n_{1} = 26$. 
Note that VAF is calculated using data from the time interval $(0.5,60]$~s.
We omit the interval $[0,0.5]$~s to reduce the impact of nonzero initial conditions.
The validation data is computed with zero initial conditions; however, the experimental data may have nonzero initial conditions.

\begin{figure}[ht!]
	\center{\includegraphics[width=0.48\textwidth,clip=true,trim= 0.14in 0.1in 0.75in 0.48in]
	{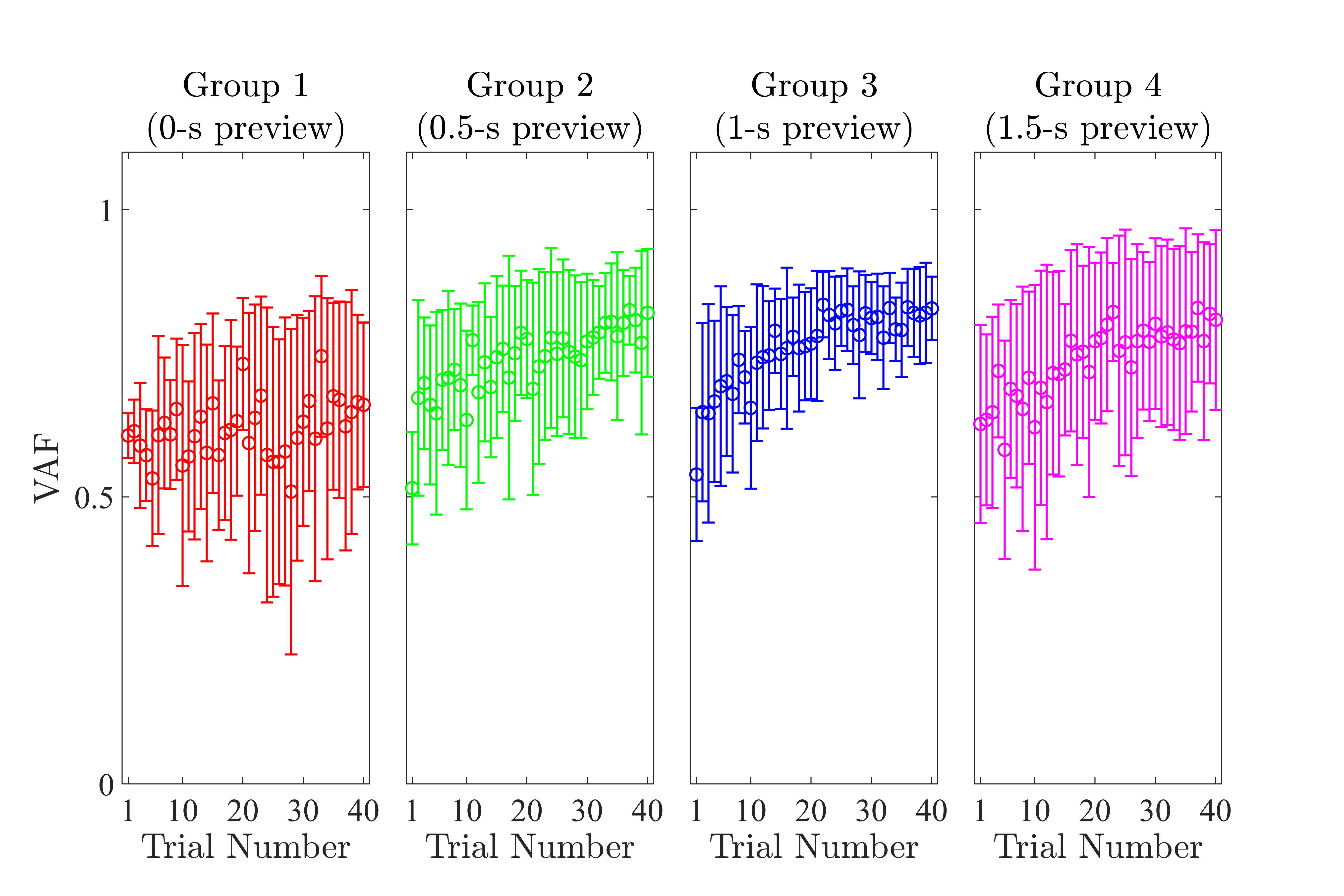}}
	\caption{
		Mean and standard deviation of VAF on each trial.
		The $\circ$ is the mean, and the lines indicate the standard deviation.
	}\label{fig:Validation}
\end{figure}

Figure~\ref{fig:Validation} shows the trial-by-trial mean and standard deviation of the VAF for each group. 
For groups 2--4, mean VAF over the last $5$ trials is greater than that over the first $5$ trials. 
Thus, as the trials progress (i.e., subjects learn), the control behavior for groups 2--4 can be more accurately modeled by the relatively low-order LTI controller \eqref{eq:ControlArchitecture} used in this study. 
These results are consistent with those in \cite{zhang2018a, zhang2022a, seyyedmousavi2020b, seyyedmousavi2020c}.


%

\vspace{-1ex}
\bibliographystyle{unsrt}
\vspace{-1ex}
\bibliography{PreviewReferences.bib}

\end{document}